\documentclass[reprint, 10pt, amsmath, amssymb, floatfix, aps, prd,  superscriptaddress, nofootinbib, nobibnotes,twocolumn]{revtex4-1}

\pdfoutput=1
\usepackage{graphicx}
\graphicspath{{figures/}}
\usepackage[utf8]{inputenc}
\usepackage{amsmath}
\usepackage{amsfonts}
\usepackage{amssymb}
\usepackage{multirow}
\usepackage{tcolorbox}
\usepackage{CJKutf8}
\usepackage{color}
\usepackage[colorlinks=true, linkcolor=red, citecolor=blue]{hyperref}
\usepackage{orcidlink}
\usepackage[capitalise]{cleveref}
\usepackage{acro}
\usepackage{adjustbox}
\usepackage{array}
\usepackage{natbib}
\usepackage{dblfnote}
\DFNalwaysdouble
\usepackage{slashed}
\usepackage{dcolumn}
\usepackage{hyperref}
\usepackage{geometry}
\usepackage{subfig} 
\usepackage{overpic}
\usepackage{inputenc}
\usepackage{caption}
\def\({\left(}
\def\){\right)}
\def\[{\left[}
\def\]{\right]}
\def\be{\begin{eqnarray}}
\def\ee{\end{eqnarray}}

\DeclareAcronym{GW}{
  short = GW ,
  long = gravitational wave ,
  short-plural = s 
}
\DeclareAcronym{LIGO}{
  short = LIGO ,
  long = Laser Interferometer Gravitational-wave Observatory ,
  short-plural = 
}
\DeclareAcronym{LISA}{
  short = LISA ,
  long = Laser Interferometer Space Antenna ,
  short-plural =  
}
\DeclareAcronym{SKA}{
  short = SKA ,
  long = Square Kilometre Array ,
  short-plural =  
}  

\DeclareAcronym{LN}{
	short = LN ,
	long = log-normal  ,
	short-plural =  
}
\DeclareAcronym{TSIGW}{
	short = TSIGW ,
	long = tensor-scalar induced gravitational wave ,
	short-plural =  s
}
\DeclareAcronym{SNR}{
	short = SNR ,
	long = signal-to-noise ratio ,
	short-plural = 
}
\DeclareAcronym{PTA}{
	short = PTA ,
	long = pulsar timing array ,
	short-plural = 
}
\DeclareAcronym{KDE}{
  short = KDE ,
  long = kernel density estimator ,
  short-plural = s
}
\DeclareAcronym{FLRW}{
  short = FLRW ,
  long = Friedmann-Lemaitre-Robertson-Walker ,
  short-plural =  
}
\DeclareAcronym{SIGW}{
	short = SIGW ,
	long = scalar induced gravitational wave ,
	short-plural =  s
}
\DeclareAcronym{PGW}{
	short = PGW ,
	long = primordial gravitational wave ,
	short-plural =  s
}
\DeclareAcronym{TIGW}{
	short = TIGW ,
	long = tensor induced gravitational wave ,
	short-plural =  s
}
\DeclareAcronym{PBH}{
	short = PBH ,
	long = primordial black hole ,
	short-plural =  s
}
\DeclareAcronym{SMBHB}{
  short = SMBHB ,
  long = supermassive black hole binary ,
  short-plural = s
}
\DeclareAcronym{CMB}{
	short = CMB ,
	long = cosmic microwave background ,
	short-plural =  
}
\DeclareAcronym{DM}{
	short = DM ,
	long = dark matter ,
	short-plural =  
}
\DeclareAcronym{BBN}{
	short = BBN ,
	long = Big-Bang nucleosynthesis ,
	short-plural =  
}
\DeclareAcronym{SGWB}{
	short = SGWB ,
	long = stochastic gravitational	wave background ,
	short-plural =  s
}
\DeclareAcronym{LSS}{
	short = LSS ,
	long = large scale structure ,
	short-plural =  
}
\DeclareAcronym{RD}{
	short = RD ,
	long = radiation-dominated ,
	short-plural =  
}
\DeclareAcronym{BAO}{
	short = BAO ,
	long = baryon acoustic oscillations ,
	short-plural = 
}

\begin{document}

\title{Probing small-scale primordial power spectra with induced gravitational waves}

\author{Di Wu\orcidlink{0000-0001-7309-574X}}
\affiliation{School of Fundamental Physics and Mathematical Sciences, Hangzhou Institute for Advanced Study, University of Chinese Academy of Sciences, Hangzhou 310024, China}

\author{Zhi-Chao Li\orcidlink{0009-0005-7984-2626}}
\affiliation{Center for Joint Quantum Studies and Department of Physics,
School of Science, Tianjin University, Tianjin 300350, China}

\author{Peng-Yu Wu\orcidlink{0009-0001-5797-3829}}
\affiliation{Center for Joint Quantum Studies and Department of Physics,
School of Science, Tianjin University, Tianjin 300350, China}

\author{Fei-Yu Chen\orcidlink{0000-0002-8674-9316}}
\affiliation{Center for Joint Quantum Studies and Department of Physics,
School of Science, Tianjin University, Tianjin 300350, China}

\author{Jing-Zhi Zhou\orcidlink{0000-0003-2792-3182}} 
\email{zhoujingzhi@tju.edu.cn}
\affiliation{Center for Joint Quantum Studies and Department of Physics,
School of Science, Tianjin University, Tianjin 300350, China}

\begin{abstract}
Large-scale primordial perturbations have been well constrained by current cosmological observations, but the properties of small-scale perturbations remain elusive. This study focuses on second-order induced gravitational waves generated by large-amplitude primordial scalar and tensor perturbations on small scales. In this case, the induced gravitational waves include contributions from three types of source terms: scalar-scalar, tensor-scalar, and tensor-tensor. To distinguish them from second-order \acp{SIGW}, we refer to those generated by these three source terms as \acp{TSIGW}. We provide the analytical expressions for the kernel functions and the corresponding energy density spectra of second-order \acp{TSIGW}. By combining observations of \ac{SGWB} across different scales, \acp{TSIGW} can be used to constrain small-scale primordial curvature perturbations and primordial gravitational waves. Furthermore, we discuss the feasibility of \acp{TSIGW} dominating the current \ac{PTA} observations under various primordial power spectra scenarios. Our results indicate that \acp{TSIGW} generated by monochromatic primordial power spectra might be more likely to dominate the current \ac{PTA} observations.
\end{abstract}

\maketitle
\acresetall

\section{Introduction}\label{sec:1.0}
The groundbreaking detection of gravitational waves by the \ac{LIGO} marked a monumental milestone in modern astrophysics \cite{LIGOScientific:2016aoc,LIGOScientific:2016vlm}. Throughout the previous decade, the study of \acp{GW} has gained substantial traction as a central theme in cosmology, astrophysics, and astronomy \cite{Annala:2017llu,LIGOScientific:2017adf,BICEP:2021xfz}. In particular, the discovery of \acp{GW} has opened a new window for cosmological research, enabling us to explore potential new physics in the evolution of the universe through observations across different \acp{GW} frequency bands, such as \ac{PTA}, \ac{LISA}, and \ac{LIGO} \cite{Reitze:2019iox,Caprini:2019egz,KAGRA:2018plz,NANOGrav:2023gor}.

In June 2023,  several international \ac{PTA} collaborations, such as NANOGrav \cite{NANOGrav:2023gor}, PPTA \cite{Reardon:2023gzh}, EPTA \cite{EPTA:2023fyk}, and CPTA \cite{Xu:2023wog}, provided evidence for the existence of a \ac{SGWB} in the nHz frequency range. For standard astrophysical scenarios, the \ac{PTA} signal is predominantly linked to \ac{SMBHB} \cite{Middleton:2020asl,NANOGrav:2020spf}. Additionally, the signal could arise from cosmological sources, such as phase transitions \cite{Fujikura:2023lkn,Addazi:2023jvg,Jiang:2023qbm,Xiao:2023dbb}, cosmic strings \cite{Ellis:2020ena,Ellis:2023tsl,Lazarides:2023ksx}, and \acp{SIGW} \cite{Vaskonen:2020lbd,DeLuca:2020agl,Franciolini:2023pbf,You:2023rmn,Zhu:2023gmx,Wang:2023ost,Wang:2023sij,Yu:2023jrs,Chang:2023vjk}. In Ref.~\cite{NANOGrav:2023hvm}, the NANOGrav collaboration investigates the possibility that \ac{SGWB} of different cosmological and astrophysical origins dominate the current \ac{PTA} observations. They provide constraints on the parameter space of various models based on the current \ac{PTA} observational data and systematically analyze Bayes factors for different models. The results indicate that the model where second-order \acp{SIGW} dominate \ac{PTA} observations has the largest Bayesian factor. Therefore, compared to other sources of \ac{SGWB}, \acp{SIGW} might be more likely to dominate current \ac{PTA} observations.

\acp{SIGW} originate from the primordial curvature perturbations generated during inflation. Specifically, the nonlinear coupling of first-order scalar perturbations and second-order tensor perturbation leads to the generation of second-order \acp{SIGW}, with primordial curvature perturbations providing the initial conditions \cite{Ananda:2006af,Baumann:2007zm,Malik:2008im}. Therefore, \acp{SIGW} depend on the characteristics of primordial curvature perturbations. On large scales ($\gtrsim$1 Mpc), the power spectrum of primordial curvature perturbations $\mathcal{P}_{\zeta}(k)$ is determined to be nearly scale-invariant through cosmological observation experiments, such as \ac{CMB},  \ac{LSS}, and \ac{BAO}, with an amplitude of $A_{\zeta} \approx 2\times 10^{-9}$ \cite{Planck:2018vyg}. Consequently, the influence of second-order \acp{SIGW} on large-scale cosmological observations is relatively minor \cite{Mollerach:2003nq,Bari:2021xvf,Saga:2015apa}.  However, on small scales ($\lesssim$1 Mpc),  primordial curvature perturbations remain weakly constrained by current observations. The large-amplitude primordial curvature perturbations on small scales can induce second-order \acp{GW} with significant observable effects, thereby influencing the \ac{SGWB} observations across various frequency bands. By combining current and future cosmological observations of different types and scales, we can constrain the physical properties of small-scale primordial perturbations \cite{Iovino:2024tyg,Perna:2024ehx,Yi:2023mbm,Wang:2025kbj,Yi:2023tdk,Yi:2023npi,Peng:2021zon,Zhao:2022kvz,Wang:2023div}.

Similar to primordial curvature perturbations, primordial tensor perturbations (primordial \acp{GW}) generated during inflation are also stringently constrained by cosmological observations on large scales. More precisely, on large scales, the tensor-to-scalar ratio $r$ is limited to $r<0.064$ at a $95\%$ confidence level \cite{Planck:2018vyg}. However, the \acp{PGW} on small scales remain relatively unconstrained. Through nonlinear coupling between first-order perturbations and second-order tensor perturbations, large-amplitude primordial tensor perturbations on small scales, in combination with primordial curvature perturbations, give rise to higher-order \acp{GW} known as \acp{TSIGW}. Here, \acp{TSIGW} are defined as the second-order \acp{GW} generated under the simultaneous presence of substantial primordial curvature perturbations and primordial tensor perturbations on small scales. They represent the combined contributions of scalar-scalar, tensor-scalar, and tensor-tensor source terms, rather than being limited to tensor-scalar contributions alone. When we neglect the potential large-amplitude primordial tensor perturbations on small scales, the second-order induced gravitational waves contain only scalar-scalar source terms. In this case, the second-order \acp{TSIGW} reduce to the well-known second-order \acp{SIGW}. In Ref.~\cite{Gong:2019mui}, the equations of motion of second-order  \acp{TSIGW} and their analytical solutions are studied comprehensively. In particular, Ref.~\cite{Chang:2022vlv} provides the first calculation of the energy density spectrum of second-order \acp{TSIGW} under a monochromatic primordial power spectrum and analyzes the impact of primordial tensor perturbations on the signal-to-noise ratio of \ac{LISA}. Subsequently, Refs.~\cite{Wu:2024qdb,Yu:2023lmo,Bari:2023rcw,Picard:2023sbz} explore second-order \acp{TSIGW} in the context of \ac{LN} primordial power spectra. When there is a significant scale disparity between primordial curvature perturbations and primordial tensor perturbations, the potential "divergence" issue of second-order \acp{SIGW} has been preliminarily studied \cite{Bari:2023rcw}. Furthermore, the cross-correlation functions of \acp{TSIGW} and the impact of primordial non-Gaussianity on \acp{TSIGW} have also been systematically analyzed in Ref.~\cite{Chen:2022dah} and Ref.~\cite{Picard:2024ekd}.

In this study, we further simplify the formula for the energy density spectrum of second-order \acp{TSIGW} and analytically derive its explicit expression. By combining current cosmological observations on different scales, we can constrain the parameter space of different forms of the primordial power spectrum and rigorously analyze the possibility that \acp{TSIGW} dominate current \ac{PTA} observations. To address potential "divergence" issues in \acp{TSIGW}, we consider the phenomenological approach proposed in Ref.~\cite{Bari:2023rcw} to study the scenario where the peak scales of primordial curvature perturbations and \acp{PGW} differ. Furthermore, within the framework of the Nieh-Yan modified Teleparallel Gravity model \cite{Fu:2023aab}, we investigate the case in which second-order \acp{GW} arise purely from large primordial tensor perturbations, without significant primordial curvature perturbations at small scales, and explore the constraints placed by current cosmological observations on small-scale primordial tensor perturbations.

This paper is organized as follows. In Sec.~\ref{sec:2.0}, we systematically revisit the key results of \acp{TSIGW} and further simplify the formula of the energy density spectrum. In Sec.~\ref{sec:3.0}, we investigate the constraints on the primordial power spectrum imposed by \ac{SGWB} and \acp{PBH}. In Sec.~\ref{sec:4.0}, we analyze the constraints imposed by current cosmological observations on various forms of the primordial power spectra, and examine scenarios in which the peaks of primordial curvature perturbations and primordial gravitational waves occur at different scales. In Sec.~\ref{sec:5.0}, we study \acp{TIGW} and the corresponding cosmological observations. Finally, we summarize our results and give some discussions in Sec.~\ref{sec:6.0}.

\section{Review of \acp{TSIGW}}\label{sec:2.0}
In this section, we briefly review the main results of second-order  \acp{TSIGW} \cite{Chang:2022vlv,Yu:2023lmo,Bari:2023rcw,Gong:2019mui,Wu:2024qdb}. The line element of perturbed spacetime in Newtonian gauge is expressed as
\cite{Malik:2008im}
\begin{eqnarray}\label{eq:dS}
	\mathrm{d} s^{2}&=&a^{2}\left[-\left(1+2 \phi^{(1)}\right) \mathrm{d} \eta^{2}+\left(\left(1-2 \psi^{(1)}\right) \delta_{i j} \right.\right.\nonumber\\
    &+&\left.\left.h^{(1)}_{ij}+\frac{1}{2}h^{(2)}_{ij}\right)\mathrm{d} x^{i} \mathrm{d} x^{j}\right] \ ,
\end{eqnarray}
where $\phi^{(1)}$ and $\psi^{(1)}$ are first-order scalar perturbations, $h^{(n)}_{ij}$$\left( n=1,2 \right)$ are $n$th-order tensor perturbations. In the study of second-order \acp{SIGW}, we only consider primordial scalar perturbations with large amplitudes on small scales, while neglecting the primordial tensor and vector perturbations. In this case, the first-order tensor perturbation $h^{(1)}_{ij}=0$ in Eq.~(\ref{eq:dS}). For \acp{TSIGW}, we need to account for both large-amplitude primordial scalar and tensor perturbations on small scales. Hence, the metric perturbations in Eq.~(\ref{eq:dS}) must include both first-order tensor and scalar perturbations. Given that vector perturbations decay proportionally to $1/a^2$, it is generally difficult to produce large primordial vector perturbations during inflation without employing a specialized inflationary model \cite{Graham:2015rva,Okano:2020uyr,Bhaumik:2025kuj}. 

\subsection{Equations of motion}\label{sec:2.1}
The equations of motion for these second-order \acp{TSIGW} can be expressed as 
\cite{Chang:2022vlv}
\begin{equation}\label{eq:h}
	\begin{aligned}
		h_{ij}^{(2)''}(\eta,\mathbf{x})&+2 \mathcal{H} h_{ij}^{(2)'}(\eta,\mathbf{x})-\Delta h_{ij}^{(2)}(\eta,\mathbf{x})=-4 \Lambda_{ij}^{lm}\\
        &\left( \mathcal{S}^{(2)}_{lm,\phi\phi}+\mathcal{S}^{(2)}_{lm,\phi h}+\mathcal{S}^{(2)}_{lm,hh}\right)  \ ,
	\end{aligned}
\end{equation}
where  $\mathcal{H}=a'/a$ is the conformal Hubble parameter, and $\Lambda_{ij}^{lm}$ is the decomposed operator to extract the transverse and traceless terms \cite{Chang:2020tji}. During the \ac{RD} era, $\mathcal{H}=1/\eta$ and $w=c_s^2=1/3$. The three types of source terms of \acp{TSIGW} in Eq.~(\ref{eq:h}) are given by
\begin{eqnarray}
	\mathcal{S}^{(2)}_{lm,\phi\phi}(\eta,\mathbf{x})&=&\partial_{l} \phi^{(1)} \partial_{m} \phi^{(1)} +4 \phi^{(1)} \partial_{l} \partial_{m} \phi^{(1)}\nonumber\\
    &-&\frac{1}{ \mathcal{H}}\left(\partial_{l} \phi^{(1)'} \partial_{m} \phi^{(1)}+\partial_{l} \phi^{(1)}  \partial_{m} \phi^{(1)'}\right) \nonumber\\
	&-&\frac{1}{ \mathcal{H}^{2}} \partial_{l} \phi^{(1)'} \partial_{m}  \phi^{(1)'} \ ,
	\label{eq:1} 
    \end{eqnarray}
\begin{eqnarray}
	\mathcal{S}^{(2)}_{lm,\phi h}(\eta,\mathbf{x})&=& 10\mathcal{H}h^{(1)}_{lm}\phi^{(1)'}+3h^{(1)}_{lm}\phi^{(1)''}\nonumber\\
&-&2\partial_{b}h^{(1)}_{lm}\partial^{b}\phi^{(1)}-2\phi^{(1)}\Delta h^{(1)}_{lm} \nonumber\\
&-&\frac{5}{3}h^{(1)}_{lm}\Delta\phi^{(1)} \ , \label{eq:2} 
\end{eqnarray}
\begin{eqnarray}
	\mathcal{S}^{(2)}_{lm,hh}(\eta,\mathbf{x})&=&\frac{1}{2}\left( -h^{b,(1)'}_{l} h^{(1)'}_{mb} +\partial_c h^{(1)}_{mb}\partial^c h^{b,(1)}_{l} \right.\nonumber\\
    &-&\left.h^{bc,(1)}\partial_c \partial_m h^{(1)}_{lb}  -\partial_b h^{(1)}_{mc}\partial^c h^{b,(1)}_{l}  \right. \nonumber\\
	&+&\left.\frac{1}{2} h^{bc,(1)}\partial_l \partial_m h^{(1)}_{bc}+ h^{bc,(1)}\partial_c \partial_b h^{(1)}_{lm}\right.\nonumber\\
    &-&\left.h^{bc,(1)}\partial_c \partial_l h^{(1)}_{mb} \right) \ .
	\label{eq:3} 
\end{eqnarray}
The source terms provided in Eq.~(\ref{eq:1})--Eq.~(\ref{eq:3}) correspond to first-order scalar source terms, source terms of first-order scalar and tensor perturbations, and first-order tensor source terms, respectively. Similar to the \acp{SIGW}, the Green's function method can be employed to solve the equations of motion for second-order \acp{TSIGW}. The solution to the second-order \acp{TSIGW} can be expressed as
\begin{equation}\label{eq:h10}
	\begin{aligned}
		h^{\lambda,(2)}(\eta,\mathbf{k})=\sum^{7}_{i=1} h_i^{\lambda,(2)}(\eta,\mathbf{k}) \ ,
	\end{aligned}
\end{equation}
where $h^{\lambda,(2)}(\eta,\mathbf{k})=\varepsilon^{\lambda, ij}(\mathbf{k})h_{ij}^{(2)}(\eta,\mathbf{k})$. The 
symbol $\varepsilon_{i j}^{\lambda}(\mathbf{k})$ represents the polarization tensor. The explicit expressions of $h_i^{\lambda,(2)}(\eta,\mathbf{k})$ $(i=1\sim 7)$ in Eq.~(\ref{eq:h10}) are given by
\begin{eqnarray}
	h_1^{\lambda,(2)}(\eta,\mathbf{k})&=&-\frac{4}{9}\int\frac{\mathrm{d}^3p}{(2\pi)^{3/2}}\varepsilon^{\lambda,lm}\left(\mathbf{k}\right)p_lp_m \nonumber\\
    &&I^{(2)}_{1}\left( u,v,x \right)\zeta_{\mathbf{k}-\mathbf{p}}\zeta_{\mathbf{p}} \ , \label{eq:h1}
\end{eqnarray}
\begin{eqnarray}
	h_2^{\lambda,(2)}(\eta,\mathbf{k})&=&-\frac{2}{3}\int\frac{\mathrm{d}^3p}{(2\pi)^{3/2}}\varepsilon^{\lambda,lm}\left(\mathbf{k}\right) \varepsilon^{\lambda_1}_{lm}\left(\mathbf{p}\right)\nonumber\\
    &&k^2 I^{(2)}_{2}\left( u,v,x \right) \zeta_{\mathbf{k}-\mathbf{p}}\mathbf{h}^{\lambda_1}_{\mathbf{p}} \ ,\label{eq:h2}
    \end{eqnarray}
\begin{eqnarray}
	h_3^{\lambda,(2)}(\eta,\mathbf{k})&=&-\int\frac{\mathrm{d}^3p}{(2\pi)^{3/2}}\varepsilon^{\lambda,lm}\left(\mathbf{k}\right) \varepsilon^{\lambda_1,b}_{l}\left(\mathbf{k}-\mathbf{p}\right)\nonumber\\
&&\varepsilon^{\lambda_2}_{bm}\left(\mathbf{p}\right)k^2I^{(2)}_{3}\left(u,v,x\right)\mathbf{h}^{\lambda_1}_{\mathbf{k}-\mathbf{p}}\mathbf{h}^{\lambda_2}_{\mathbf{p}} \ ,\label{eq:h3}
\end{eqnarray}
\begin{eqnarray}
	h_4^{\lambda,(2)}(\eta,\mathbf{k})&=&-\int\frac{\mathrm{d}^3p}{(2\pi)^{3/2}}\varepsilon^{\lambda,lm}\left(\mathbf{k}\right)\varepsilon^{\lambda_2}_{mb}\left( \mathbf{p} \right)\nonumber\\
 &&2\varepsilon^{\lambda_1,bc}\left( \mathbf{k}-\mathbf{p} \right)p_cp_lI^{(2)}_4\left(u,v,x\right)\nonumber\\ &&\mathbf{h}^{\lambda_1}_{\mathbf{k}-\mathbf{p}}\mathbf{h}^{\lambda_2}_{\mathbf{p}} \ ,\label{eq:h4}
 \end{eqnarray}
\begin{eqnarray}
	h_5^{\lambda,(2)}(\eta,\mathbf{k})&=&-\int\frac{\mathrm{d}^3p}{(2\pi)^{3/2}}\varepsilon^{\lambda,lm}\left(\mathbf{k}\right)\left( k-p \right)^bp^c \nonumber\\
&&\varepsilon^{\lambda_1}_{mc}\left(\mathbf{k}-\mathbf{p}\right)\varepsilon^{\lambda_2}_{lb}\left(\mathbf{p}\right)I^{(2)}_{5}\left(u,v,x\right) \nonumber\\
 & & \mathbf{h}^{\lambda_1}_{\mathbf{k}-\mathbf{p}}\mathbf{h}^{\lambda_2}_{\mathbf{p}} \ ,\label{eq:h5}
 \end{eqnarray}
\begin{eqnarray}
	h_6^{\lambda,(2)}(\eta,\mathbf{k})&=&\int\frac{\mathrm{d}^3p}{(2\pi)^{3/2}}\varepsilon^{\lambda,lm}\left(\mathbf{k}\right) \varepsilon^{\lambda_1}_{bc}\left(\mathbf{k}-\mathbf{p}\right)\varepsilon^{\lambda_2}_{lm}\left(\mathbf{p}\right)\nonumber\\
&&p^bp^cI^{(2)}_6\left(u,v,x\right)\mathbf{h}^{\lambda_1}_{\mathbf{k}-\mathbf{p}}\mathbf{h}^{\lambda_2}_{\mathbf{p}} \ ,\label{eq:h6} 
\end{eqnarray}
\begin{eqnarray}
	h_7^{\lambda,(2)}(\eta,\mathbf{k})&=&\int\frac{\mathrm{d}^3p}{(2\pi)^{3/2}}\varepsilon^{\lambda,lm}\left(\mathbf{k}\right) \varepsilon^{\lambda_1,bc}\left(\mathbf{k}-\mathbf{p}\right)\nonumber\\
&&\varepsilon^{\lambda_2}_{bc}\left(\mathbf{p}\right)\frac{p_lp_m}{2} I^{(2)}_{7}\left(u,v,x\right) \nonumber\\
&&\mathbf{h}^{\lambda_1}_{\mathbf{k}-\mathbf{p}}\mathbf{h}^{\lambda_2}_{\mathbf{p}} \ .	\label{eq:h7}
\end{eqnarray}
Here, we have defined $|\mathbf{k}-\mathbf{p}|=u|\mathbf{k}|$ and $|\mathbf{p}|=v|\mathbf{k}|$. In Eq.~(\ref{eq:h10}), based on the specific form of the source term, we have decomposed the contributions of the source term  $\mathcal{S}^{(2)}_{lm,hh}$ into five parts: $h^{\lambda,(2)}_3 \sim h^{\lambda,(2)}_7$. The analytic expressions of the kernel functions $I^{(2)}_i\left( u,v,x \right)$ in Eq.~(\ref{eq:h1})--Eq.~(\ref{eq:h7}) are given by
\begin{eqnarray}
        I^{(2)}_1&=&
        \frac{27 (u^2 + v^2 - 3)}{k^2u^3 v^3 x}
        \bigg(
            \sin x 
            \left(
                -4 u v + (u^2  \right.\nonumber\\
            &+&\left.v^2- 3) 
                \ln \left| \frac{3 - (u + v)^2}{3 - (u - v)^2} \right|
            \right)-\pi(u^2 \nonumber\\
        &+& v^2 - 3) \Theta(v + u - \sqrt{3}) \cos x
        \bigg) \ , \label{eq:11I}
\end{eqnarray}
\begin{eqnarray}
I^{(2)}_2&=&
        \frac{\sqrt{3} \left(u^2 - 3 (1-v)^2\right)}{16k^2 u^3 v x}\nonumber\\
        &\times&\Bigg[
            \sin x 
            \bigg(
                \left(u^2 - 3 (1+v)^2\right) 
                \ln \left|
                    \frac{\left(u - \sqrt{3} v\right)^2 - 3}{\left(u + \sqrt{3} v\right)^2 - 3}
                \right| \nonumber\\
         &-&\frac{4 u v \left(u^2 - 9 v^2 + 9\right)}{\sqrt{3} \left(u^2 - 3 (1-v)^2\right)}
            \bigg) + \pi \left(u^2 - 3 (1+v)^2\right) \nonumber\\
        &\times&\Theta (u -  \sqrt{3}|1 - v|) \cos x 
        \Bigg]\ , \label{eq:22I}
\end{eqnarray}
\begin{eqnarray}
        I^{(2)}_3&=&\frac{1}{k^2}\left(
        \frac{\sin x}{4 x} 
        - \frac{\sin(u x) \sin(v x)}{4 u v x^2}\right)\ , \label{eq:33I}\\
        I^{(2)}_i&=&\frac{1}{8k^2uvx}\Bigg[\sin x \times\ln\left|\frac{1 - (u + v)^2}{1 - (u - v)^2}\right|  \nonumber \\
        &-& \cos x \times\pi\times \Theta\left(u-|1-v|\right)
        \Bigg] \  , \ (i=4,5,6,7) \ ,\nonumber\\ 
        \label{eq:ijI}
\end{eqnarray}
where $\Theta(x)$ represents the Heaviside theta function. We have used the following approximations: $\lim_{x\to\pm \infty} \mathrm{Si}(x)=\pm \pi/2$ and $\lim_{x\to \infty} \mathrm{Ci}(x)=0$ in Eq.~(\ref{eq:11I}) and Eq.~(\ref{eq:ijI}). 

\subsection{Energy density spectra}\label{sec:2.2}
To obtain the energy density spectrum of second-order \acp{TSIGW}, it is necessary to employ Eq.~(\ref{eq:h1})--Eq.~(\ref{eq:h7}) to calculate the corresponding two-point correlation function. More precisely, the total energy density fraction of \acp{GW} up to second order can be written as \cite{Maggiore:1999vm}
\begin{equation}\label{eq:Omega}
	\begin{aligned}
		\Omega_{\mathrm{GW}}^{\mathrm{tot}}(\eta, k)&=\Omega_{\mathrm{GW}}^{(1)}(\eta, k)+\Omega_{\mathrm{GW}}^{(2)}(\eta, k)\\
		&=\frac{1}{6}\left(\frac{k}{a(\eta) H(\eta)}\right)^{2} \mathcal{P}^{\mathrm{tot}}_{h}(\eta, k) \ ,
	\end{aligned}
\end{equation}
where
\begin{eqnarray}\label{eq:P12}
    \mathcal{P}^{\mathrm{tot}}_{h}(\eta, k)=\mathcal{P}^{(1)}_{h}(\eta, k) +\frac{1}{4}\mathcal{P}^{(2)}_{h}(\eta, k) \ .
\end{eqnarray}
In Eq.~(\ref{eq:P12}), $\mathcal{P}^{(1)}_{h}(\eta, k)$ and $\mathcal{P}^{(2)}_{h}(\eta, k)$ represent the power spectra of first-order \acp{GW} and second-order \acp{TSIGW}, respectively. The power spectra of $n$-th order \acp{GW} $\mathcal{P}_{h}^{(n)}( \mathbf{k},\eta)$ in Eq.~(\ref{eq:P12}) are defined as
\begin{equation}\label{eq:Ph}
  \left\langle h^{\lambda,(n)}_{\mathbf{k}}h^{\lambda^{\prime},(n)}_{\mathbf{k}^{\prime}}\right\rangle=\delta^{\lambda \lambda^{\prime}} \delta\left(\mathbf{k}+\mathbf{k}^{\prime}\right) \frac{2 \pi^2}{k^3} \mathcal{P}_{h}^{(n)}(k,\eta) \ .  
\end{equation}
Here, we assume that the two-point correlation function between the primordial curvature perturbations and primordial tensor perturbations $\left\langle \zeta_{\mathbf{k}} \mathbf{h}^{\lambda}_{\mathbf{k}}\right\rangle=0$. By calculating the two-point correlation functions of gravitational waves in Eq.~(\ref{eq:h1})--Eq.~(\ref{eq:h7}), we can use Eq.~(\ref{eq:Omega}) and Eq.~(\ref{eq:P12}) to derive the total energy density spectrum of \acp{GW}. 

In this study, we further simplify the formula for calculating the energy density spectrum of \acp{TSIGW}. The explicit expression for the energy density spectrum of  \acp{TSIGW} is given by
\begin{eqnarray}\label{eq:O3}
    \Omega_{\mathrm{GW}}^{(2)}(k)=\Omega_{\mathrm{GW}}^{ss}(k)+\Omega_{\mathrm{GW}}^{st}(k)+\Omega_{\mathrm{GW}}^{tt}(k) \ .
\end{eqnarray}
In Eq.~(\ref{eq:O3}), $\Omega_{\mathrm{GW}}^{ss}(k)$, $\Omega_{\mathrm{GW}}^{st}(k)$, and $\Omega_{\mathrm{GW}}^{tt}(k)$ represent the energy density spectra induced by the scalar-scalar source, tensor-scalar source, and tensor-tensor source, respectively. The specific expressions for the three energy density spectra are given by
\begin{align}
    \Omega_{\mathrm{GW}}^{ss}(k) &= \int_{0}^{\infty} \mathrm{d}v \int_{|1-v|}^{1+v} \mathrm{d}u\,
    \mathcal{P}_{\zeta}(uk) \mathcal{P}_{\zeta}(vk) \notag \\
    &\times  \frac{3}{1024 u^8 v^8} (u^2 + v^2 - 3)^2 \notag \\
    &\times \left[4v^2 - (1 + v^2 - u^2)^2\right]^2 \notag \\
    &\times \left\{ \left[(u^2 + v^2 - 3) \ln \left| \frac{3 - (u + v)^2}{3 - (u - v)^2} \right| - 4uv \right]^2 \right. \notag \\
    & + \left.  \pi^2 (u^2 + v^2 - 3)^2 \Theta \Big(u + v - \sqrt{3} \Big) \right\} \,,
    \label{eq:1Oss}
\end{align}
\begin{equation}
    \begin{aligned}
        \Omega_{\mathrm{GW}}^{st}(k)&=\int_{0}^{\infty} \mathrm{d}v \int_{|1-v|}^{1+v} \mathrm{d}u\, \mathcal{P}_{\zeta}(uk) \mathcal{P}_h(vk)  \\
        &\times \frac{1}{442368 u^8 v^8} \bigg[ 16v^4 + 24v^2 (1 + v^2 - u^2)^2  \\ 
        & + (1 + v^2 - u^2)^4 \bigg] \Bigg\{
        \Bigg[ 4 u v \Big(u^2 - 9(v^2 - 1)\Big) \\
        &+\sqrt{3}\Big(u^2 - 3(v - 1)^2 \big) \Big(u^2 - 3(v + 1)^2 \Big) \\
        & \times \ln \left| \frac{3 - (u + \sqrt{3}v)^2}{3 - (u - \sqrt{3}v)^2} \right| \Bigg]^2 + \Theta\Big(u^2 - 3(v - 1)^2\Big)\\
        & \times 3\pi^2 \Big(u^2 - 3(v - 1)^2 \Big)^2 \Big(u^2 - 3(v + 1)^2 \Big)^2  \Bigg\} \ , \label{eq:1Ost}
    \end{aligned}
\end{equation}
\begin{equation}
    \begin{aligned}
        \Omega_{\mathrm{GW}}^{tt}(k)&= \int_{0}^{\infty} \mathrm{d}v \int_{|1-v|}^{1+v} \mathrm{d}u\, 
        \mathcal{P}_h(uk) \mathcal{P}_h(vk) \\
        &\times \frac{1}{3145728\, u^8 v^8} 
        \Big((u - v)^2 - 1\Big)^2 \Big((u + v)^2 - 1\Big)^2 \\
        &\times \Bigg[
        64 u^2 v^2 \Big(1 + u^4 + v^4 + 6 (u^2 + v^2) + 6 u^2 v^2\Big) \\
        & + 16 u v \Big(1 + u^6 + v^6 + 15 (u^4 + v^4) + 15 (u^2 + v^2) \\
        &\quad + 15 u^2 v^2 (u^2 + v^2 + 6) \Big)
        \ln \left| \frac{1 - (u + v)^2}{1 - (u - v)^2} \right| \\
        &\quad + \Big(
            1 - 7 u^8 + 4 v^2 + 126 v^4 + 116 v^6 + 9 v^8 \\
        &\quad - 12 u^6 (5 + 7 v^2)
            + 2 u^4 (7 + 118 v^2 + 35 v^4) \\
        &\quad + 4 u^2 (13 + 105 v^2 + 151 v^4 + 35 v^6)
        \Big) \\
        &\quad \times \left( \pi^2 + \ln^2 \left| \frac{1 - (u + v)^2}{1 - (u - v)^2} \right| \right)
        \Bigg] \ ,
    \end{aligned}
    \label{eq:1Ott}
\end{equation}
where $\mathcal{P}_{\zeta}(k)$ and $\mathcal{P}_{h}(k)$ represent the primordial power spectra of the curvature perturbation and tensor perturbation, respectively. {It should be noted that Eq.~(\ref{eq:1Oss}) provides the formula for calculating the energy density spectrum of second-order \acp{SIGW}, and the related results have been systematically studied in Refs.~\cite{Kohri:2018awv,Espinosa:2018eve}. Furthermore, Eq.~(\ref{eq:1Ost}) and Eq.~(\ref{eq:1Ott}) correspond to the explicit expressions of the energy density spectra of second-order \acp{GW} generated by tensor-scalar and tensor-tensor source terms in second-order \acp{TSIGW}, respectively. Similar results have also been presented in Refs.~\cite{Chang:2022vlv,Picard:2023sbz}. However, in previous works, the formulas for the energy density spectra were not systematically simplified. For example, Eq.~(39) in Ref.~\cite{Chang:2022vlv} gives the expression for the energy density spectrum corresponding to the tensor-tensor source term. Nevertheless, this result involves complicated momentum polynomial calculations, making it cumbersome for direct application in energy density spectrum calculations. In this study, following the methodology applied to second-order \acp{SIGW}, we systematically simplify the formulas for the remaining two spectra of \acp{TSIGW} and derive compact expressions suitable for direct computation.} All calculations related to second-order \acp{TSIGW} are encapsulated in Eq.~(\ref{eq:1Oss})--Eq.~(\ref{eq:1Ott}). For any inflation model, by substituting the primordial power spectra derived into Eq.~(\ref{eq:1Oss})--Eq.~(\ref{eq:1Ott}), the corresponding energy density spectrum of second-order \acp{TSIGW} can be obtained. 

Eq.~(\ref{eq:Omega}) provides the energy density spectrum of \acp{PGW}$+$\acp{TSIGW} during the \ac{RD} era. Taking into account the thermal history of the universe, the current total energy density spectrum $\bar{\Omega}^{(2)}_{\mathrm{GW,0}}(k)$ is given by
\begin{equation}
    \bar{\Omega}^{\mathrm{tot}}_{\mathrm{GW,0}}(k) = \Omega_{\mathrm{rad},0}\left(\frac{g_{*,\rho,e}}{g_{*,\rho,0}}\right)\left(\frac{g_{*,s,0}}{g_{*,s,e}}\right)^{4/3}\bar{\Omega}^{\mathrm{tot}}_{\mathrm{GW}}(k) \ ,
\end{equation}
where $\Omega_{\mathrm{rad},0}$ ($ =4.2\times 10^{-5}h^{-2}$) is the energy density fraction of radiation today.

\section{\ac{SGWB} and \ac{PBH}}\label{sec:3.0}
In Sec.~\ref{sec:2.0}, we present the explicit expression for the energy density spectrum of \acp{TSIGW}. As shown in Eq.~(\ref{eq:O3}), the energy density spectrum of \acp{TSIGW} consists of three contributions: spectrum $\Omega_{\mathrm{GW}}^{ss}(k)$ generated by scalar-scalar source terms, spectrum $\Omega_{\mathrm{GW}}^{st}(k)$ generated by scalar-tensor source terms, and spectrum $\Omega_{\mathrm{GW}}^{tt}(k)$ generated by tensor-tensor source terms. In this section, we will compute the total energy density spectrum of second-order \acp{TSIGW} under different forms of primordial power spectra and analyze the impact of the three distinct source contributions on the overall spectrum. Furthermore, we will systematically analyze the impact of small-scale primordial power spectra on current cosmological observations. By integrating current \ac{PTA}, \ac{BAO}, \ac{CMB}, and \ac{PBH} observational data, we can examine the constraints imposed by current cosmological observations on the parameter space of small-scale primordial power spectra. 

\subsection{Observations of \ac{SGWB}}\label{sec:3.1}
On small scales, potential primordial perturbations with large amplitudes serve as initial conditions that influence subsequent cosmological evolution. Through current small-scale cosmological observations, we can constrain the physical properties of the primordial power spectrum on small scales. For the \ac{SGWB} observations, we focus primarily on cases where \acp{TSIGW} dominate the current observations of the \ac{SGWB}. By combining \ac{PTA} and \ac{LISA} observations of the energy density spectrum of the \ac{SGWB} and large-scale cosmological constraints on its energy density, we can constrain the parameter space of various small-scale primordial power spectra. 

\subsubsection{\ac{PTA} observation}
For \ac{PTA} observation, we employ the \ac{KDE} representations of the free spectra to construct the likelihood function \cite{Mitridate:2023oar,Lamb:2023jls,Moore:2021ibq}
\begin{equation} \label{eq:likelihood}
    \ln \mathcal{L}(d|\theta) = \sum_{i=1}^{N_f} p(\Phi_i,\theta)\ .
\end{equation}
In Eq.~(\ref{eq:likelihood}), $p(\Phi_i,\theta)$ represents the probability of $\Phi_i$ given the parameter $\theta$, and $\Phi_i = \Phi(f_i)$ denotes the time delay
\begin{equation} \label{eq:timedelay}
    \Phi(f) = \sqrt{\frac{H_0^2 \Omega_{\mathrm{GW}}(f)}{8\pi^2 f^5 T_{\mathrm{obs}}}} \ ,
\end{equation}
where $H_0=h\times 100 \mathrm{km/s/Mpc}$ is the present-day value of the Hubble constant. In this study, we utilize the \acp{KDE} representation of the first 14 frequency bins of the HD-correlated free spectrum from the NANOGrav 15-year dataset \cite{Nanograv:KDE}. The Bayesian analysis is performed via \textsc{bilby} \cite{bilby_paper} using its integrated \textsc{dynesty} nested sampler \cite{Speagle:2019ivv,dynesty_software}. Furthermore, to rigorously assess the feasibility of different models dominating the current \ac{PTA} observations, we also consider the scenario in which both \acp{SMBHB} and \acp{TSIGW} jointly influence the present \ac{PTA} data. The energy density spectrum of \acp{SMBHB} is given by \cite{Mitridate:2023oar,NANOGrav:2023hvm}
\begin{equation} \label{eq:SMBHB}
    \Omega_{\mathrm{GW}}^{\mathrm{BH}}(f) = \frac{2\pi^2 A_{\mathrm{BHB}}^2}{3H_0^2 h^2} (\frac{f}{\mathrm{year}^{-1}})^{5-\gamma_{\mathrm{BHB}}}\mathrm{year}^{-2} \ .
\end{equation}
Here, the prior distribution for $(\log_{10}A_{\mathrm{BHB}}, \gamma_{\mathrm{BHB}})$ follows a multivariate normal distribution \cite{NANOGrav:2023hvm}, whose mean and covariance matrix are given by
\begin{equation} \label{eq:prior_SMBHB}
\begin{aligned}
    \boldsymbol{\mu}_{\mathrm{BHB}}&=\begin{pmatrix} -15.6
 \\ 4.7 \end{pmatrix} \ , \\
\boldsymbol{\sigma}_{\mathrm{BHB}}&=0.1\times \begin{pmatrix}
2.8  & -0.026\\
-0.026  & 2.8
\end{pmatrix} \ .
\end{aligned}
\end{equation} 
In addition, we employ the Bayes factor to compare different models. The Bayes factor is defined as $B_{i,j} = \frac{Z_i}{Z_j}$, where $Z_i$ represents the evidence of model $H_i$.

\subsubsection{\ac{LISA} observation}
In addition to PTA observations, we can explore \ac{SGWB} detection across different frequency bands. If \acp{TSIGW} or \acp{SMBHB} dominate the current \ac{PTA} observations, the high-frequency region of the associated energy density spectrum can be examined for its impact on \ac{LISA} by evaluating its \ac{SNR}. The \ac{SNR} of \ac{LISA} can be expressed as \cite{Siemens:2013zla, Robson:2018ifk}
\begin{equation}
    \rho = \sqrt{T}\left[ \int \mathrm{d} f\left(\frac{\bar{\Omega}_{\mathrm{GW},0}(f)}{\Omega_\mathrm{n}(f)}\right)^2\right]^{1/2} \ ,
\end{equation}
where $T$ is the observation time and we set $T=4$ years here. $\Omega_\mathrm{n}(f)=2\pi^2f^3S_n/3H_0^2$, where $H_0$ is the Hubble constant, $S_n$ is the strain noise power spectral density \cite{Robson:2018ifk}.

\subsubsection{Large-scale cosmological observations }\label{sec:3.1.3}
In addition to the direct detection of the energy density spectrum of \acp{TSIGW}, the \acp{TSIGW} can act as an extra radiation component, influencing large-scale cosmological observations \cite{Zhou:2024yke,Wright:2024awr,Ben-Dayan:2019gll,Wang:2025qpj}. Specifically, the total energy density of \acp{SIGW} must satisfy
\begin{equation}\label{eq:rhup}
h^2\rho_{\mathrm{GW}}=\int_{f_{\mathrm{min}}}^{\infty} h^2\Omega_{\mathrm{GW},0}(k) \mathrm{d}\left(\ln k\right) < 2.9\times 10^{-7}  \ ,
\end{equation}
at $95\%$ confidence level for \ac{CMB}$+$\ac{BAO} data \cite{Clarke:2020bil}. It should be noted that the results we adopt from Ref.~\cite{Clarke:2020bil} are stronger than the constraints on the energy density spectrum of \ac{SGWB} derived from $N_{\mathrm{eff}}$. Using the constraint in Eq.~(\ref{eq:rhup}), we can constrain the parameter space of primordial power spectra.

\subsection{Abundance of \acp{PBH}}\label{sec:3.2}
Small-scale primordial curvature perturbations with substantial amplitudes create notable density perturbations that drive \ac{PBH} formation. Observational constraints on the upper limits of primordial black hole abundance enable us to infer the physical characteristics of primordial curvature perturbation on small scales. Furthermore, the formation of \acp{PBH} is not directly affected by primordial tensor perturbation. It appears that we cannot use the abundance of \acp{PBH} to constrain small-scale primordial tensor perturbations. However, as pointed out in Refs.~\cite{Nakama:2015nea,Nakama:2016enz}, significant primordial tensor perturbations at small scales can induce the second-order density perturbation $\rho^{(2)}$, thereby influencing the formation of \acp{PBH}. The abundance of \ac{PBH} also acts as a constraint on small-scale tensor perturbations. 

In this study, we neglect the impact of higher-order density perturbations induced by primordial tensor perturbations \cite{DeLuca:2023tun,Zhou:2023itl}, focusing solely on the relationship between primordial curvature perturbations and \acp{PBH}. In this context, the observational constraints on \ac{PBH} abundance apply exclusively to primordial curvature perturbations, with no impact on primordial tensor perturbations. When accounting for the second-order density perturbations induced by primordial tensor perturbations in primordial black hole formation, the resulting upper bound on the amplitude of primordial curvature perturbation is reduced.

We focus on two distinct models for calculating the abundance of \acp{PBH}: the threshold statistics and the theory of peaks. It is well established that the threshold statistics does not align with the theory of peaks \cite{Balaji:2023ehk,Green:2004wb,Young:2014ana}. Therefore, we will compare the constraints imposed by these models on different parameter spaces of the primordial power spectrum.

\subsubsection{Threshold statistics}
In threshold statistics, the \ac{PBH} abundance is computed by integrating the probability distribution function of the smoothed density contrast from the threshold value $\mathcal{C}_{\mathrm{th}}$, defined as
\begin{eqnarray}\label{eq:betat}
    \beta_{\mathrm{th}}=\int_{\mathcal{D}} \mathcal{K}\left(\mathcal{C}-\mathcal{C}_{\mathrm{th}}\right)^\gamma \mathrm{P}_{\mathrm{G}}\left(\mathcal{C}_{\mathrm{G}}, \zeta_{\mathrm{G}}\right) \mathrm{d} \mathcal{C}_{\mathrm{G}} \mathrm{~d} \zeta_{\mathrm{G}} \ ,
\end{eqnarray}
where the domain of integration in Eq.~(\ref{eq:betat}) is $\mathcal{D}=$ $\left\{\mathcal{C}\left(\mathcal{C}_{\mathrm{G}}, \zeta_{\mathrm{G}}\right)>\mathcal{C}_{\text {th}} \wedge \mathcal{C}_1\left(\mathcal{C}_{\mathrm{G}}, \zeta_{\mathrm{G}}\right)<2 \Phi\right\}$. Here, we set $\mathcal{K}=4.4$ and $\gamma=0.38$ \cite{Musco:2023dak,Iovino:2024tyg}. In Eq.~(\ref{eq:betat}), the compaction function $\mathcal{C}=\mathcal{C}_1-\mathcal{C}_1^2 /(4 \Phi)$ can be obtained from the linear $\mathcal{C}_1=\mathcal{C}_{\mathrm{G}} \mathrm{d} F / \mathrm{d} \zeta_{\mathrm{G}}$ component, that uses $\mathcal{C}_{\mathrm{G}}=$ $-2 \Phi r \zeta_G^{\prime}$, where $\Phi=3(1+w)/(5+3w)$ and $\zeta=F(\zeta_{G})$. And the Gaussian components are distributed as
\begin{equation}\label{eq:PG1}
    P_{\mathrm{G}}\left(\mathcal{C}_{\mathrm{G}}, \zeta_{\mathrm{G}}\right)=\frac{e^{\left[-\frac{1}{2\left(1-\gamma_{c r}^2\right)}\left(\frac{\mathcal{C}_{\mathrm{G}}}{\sigma_c}-\frac{\gamma_{c r} \zeta_{\mathrm{G}}}{\sigma_r}\right)^2-\frac{\zeta_\mathrm{G}^2}{2 \sigma_r^2}\right]}}{2 \pi \sigma_c \sigma_r \sqrt{1-\gamma_{c r}^2}} \ ,
\end{equation}
where the correlators in Eq.~(\ref{eq:PG1}) are given by
\begin{equation}\label{eq:sgm3}
\begin{aligned}
& \sigma_c^2=\frac{4 \Phi^2}{9} \int_0^{\infty} \frac{\mathrm{d} k}{k}\left(k r_m\right)^4 W^2\left(k, r_m\right) P_\zeta^T \ , \\
& \sigma_{c r}^2=\frac{2 \Phi}{3} \int_0^{\infty} \frac{\mathrm{d} k}{k}\left(k r_m\right)^2 W\left(k, r_m\right) W_s\left(k, r_m\right) P_\zeta^T \ , \\
& \sigma_r^2=\int_0^{\infty} \frac{\mathrm{d} k}{k} W_s^2\left(k, r_m\right) P_\zeta^T \ ,
\end{aligned}
\end{equation}
where $P_\zeta^T=T^2\left(k, r_m\right) P_\zeta(k)$, and $\gamma_{c r} \equiv \sigma_{c r}^2 / \sigma_c \sigma_r$. Here, we have defined $W\left(k, r_m\right), W_s\left(k, r_m\right)$ and $T\left(k, r_m\right)$ as the top-hat window function, the spherical-shell window function, and the radiation transfer function \cite{Young:2022phe}. 

When considering Gaussian primordial curvature perturbations, we have $F(\zeta_G)=\zeta_G$. In this case, $\mathcal{C}=\mathcal{C}_{\mathrm{G}}-\mathcal{C}_{\mathrm{G}}^2 /(4 \Phi)$. And the abundance of \acp{PBH} can be expressed as \cite{Sasaki:2018dmp}
\begin{equation}\label{eq:pbh1}
\begin{aligned}
 f_{\mathrm{pbh}} \simeq 2.5 \times 10^8 \beta\left(\frac{g_*^{\text {form }}}{10.75}\right)^{-\frac{1}{4}}\left(\frac{m_{\mathrm{pbh}}}{M_{\odot}}\right)^{-\frac{1}{2}} \ .
\end{aligned}
\end{equation}
By substituting the results of Eq.~(\ref{eq:betat}) into Eq.~(\ref{eq:pbh1}), we can calculate the \ac{PBH} abundance associated with various primordial power spectra.

\subsubsection{Theory of peaks}
Unlike threshold statistics, in peak theory,  the total fraction of \acp{PBH} can be expressed as \cite{Yoo:2018kvb}
\begin{equation}\label{eq:peak1}
\beta_{\text {peaks }}=\int_{\delta_{c, l-}}^{\frac{4}{3}} \mathrm{~d} \delta_l \mathcal{K}\left(\delta_m-\delta_{\text {th }}\right)^\gamma \mathcal{N}(\nu) \ ,
\end{equation}
where $\delta_{c, l-}=\frac{4}{3}\left(1-\sqrt{(2-3 \delta_{\mathrm{th}})/2}\right)$, $\nu=\delta_l / \sigma$, and $\delta_m=\delta_l-\frac{3}{8} \delta_l^2$ \cite{Balaji:2023ehk}. In Eq.~(\ref{eq:peak1}), the number density of peaks is given by
\begin{equation}
    \mathcal{N}(\nu)=\frac{1}{3^{3 / 2} 4 \pi^2}\left(\frac{\mu}{\sigma}\right)^3 \nu^3 \exp \left(-\frac{\nu^2}{2}\right)\ ,
\end{equation}
where 
\begin{equation}
    \mu^2=\frac{4\Phi^2}{9} \int_0^{\infty} \frac{\mathrm{d} k}{k}\left(k r_m\right)^6 W^2\left(k r_m\right)  \mathcal{P}^{T}_{\zeta} \ .
\end{equation}
By applying peak theory and leveraging the results from Eq.~(\ref{eq:peak1}), we can evaluate the \ac{PBH} abundance associated with a given form of the power spectrum of primordial curvature perturbations.

\section{Constrains on small-scale primordial power spectra}\label{sec:4.0}
In this section, we apply the results from Sec.~\ref{sec:3.0} to specific forms of the primordial power spectra to constrain the parameter space of different types of primordial power spectra. Furthermore, following the phenomenological approach outlined in Ref.~\cite{Bari:2023rcw}, we conducted a preliminary investigation into scenarios where the peak scales of primordial curvature perturbations and \acp{PGW} differ.

\subsection{Monochromatic spectra}\label{sec:4.1}
\begin{figure}[htbp]
\centering
\includegraphics[width=\linewidth]{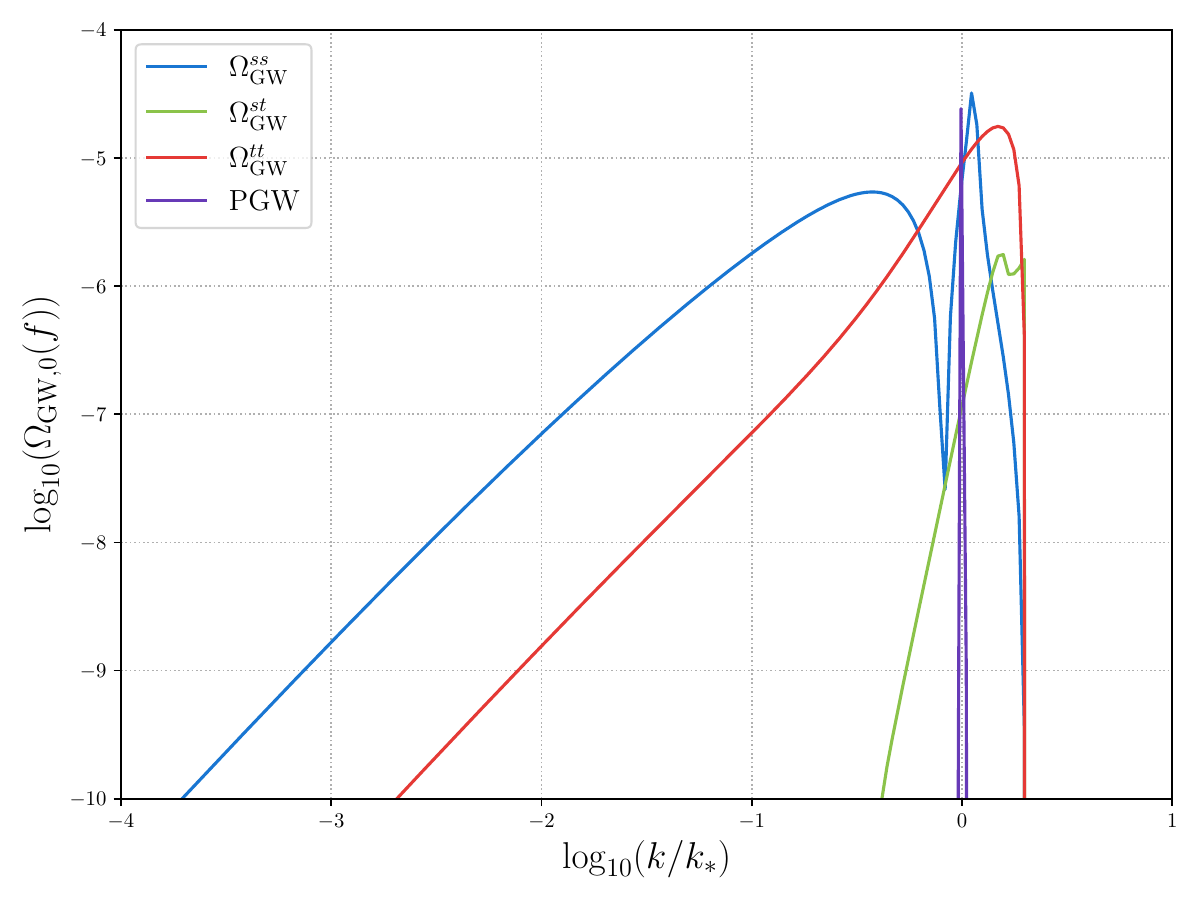}
\caption{{In the case of the monochromatic primordial power spectrum, the current energy density spectra of the three components of second-order \acp{TSIGW} are shown as blue, green, and red curves, respectively. The corresponding energy density spectrum of \acp{PGW} is represented by the purple curve. The parameters for the curves are $A_\zeta=1$ and $A_h=1$. To facilitate visualization, the \ac{LN}  power spectrum of \acp{PGW} with $\sigma = 0.01$ is adopted in the figure as a substitute for the monochromatic primordial gravitational wave power spectrum. } } \label{fig:spectrum_mono}
\end{figure}

\begin{figure}[htbp]
\centering
\includegraphics[width=\linewidth]{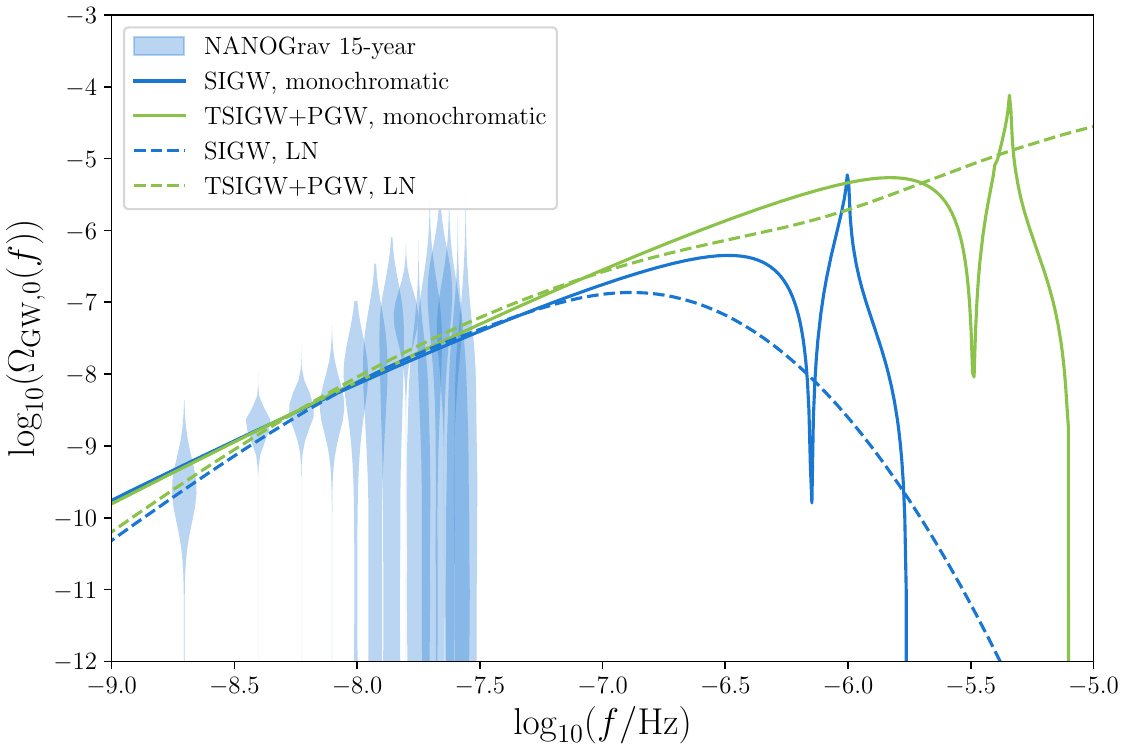}
\caption{The current energy density spectra of \acp{SIGW} and \acp{PGW}$+$\acp{TSIGW} { for monochromatic spectra and \ac{LN} spectra}. The energy density spectra derived from the free spectrum of the NANOGrav 15-year are shown in blue. The blue and green curves represent the energy density spectra of \acp{GW} with different line styles labeled in the figure. These parameters are selected based on the median values of the posterior distributions, with the median values shown as green numbers in Fig.~\ref{fig:corner_SIGW_mono}, Fig.~\ref{fig:corner_TSIGW_mono}, Fig.~\ref{fig:corner_SIGW_logn} and Fig.~\ref{fig:corner_TSIGW_logn}.} \label{fig:violinplot}
\end{figure}

\begin{figure}[htbp]
\centering
\includegraphics[width=\linewidth]{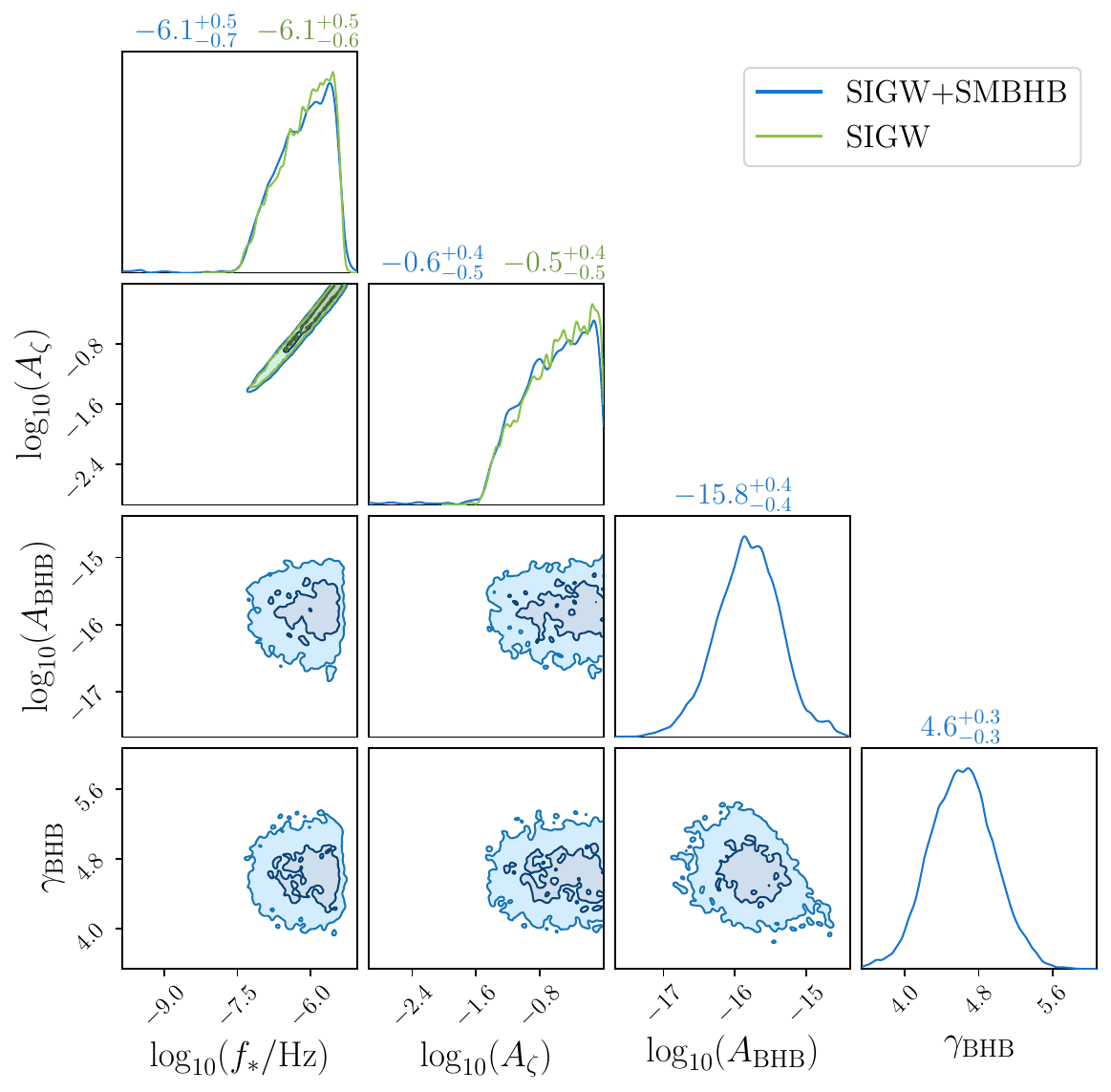}
\caption{The corner plot of the posterior distributions. The contours in the off-diagonal panels denote the $68\% $ and $95 \%$ credible intervals of the 2D posteriors. The numbers above the figures represent the median values and $1$-$\sigma$ ranges of the parameters. The blue and green solid curves correspond to the \acp{SIGW} energy spectrum with or without \ac{SMBHB}, assuming the monochromatic primordial power spectrum.} \label{fig:corner_SIGW_mono}
\end{figure}

\begin{figure}[htbp]
\centering
\includegraphics[width=\linewidth]{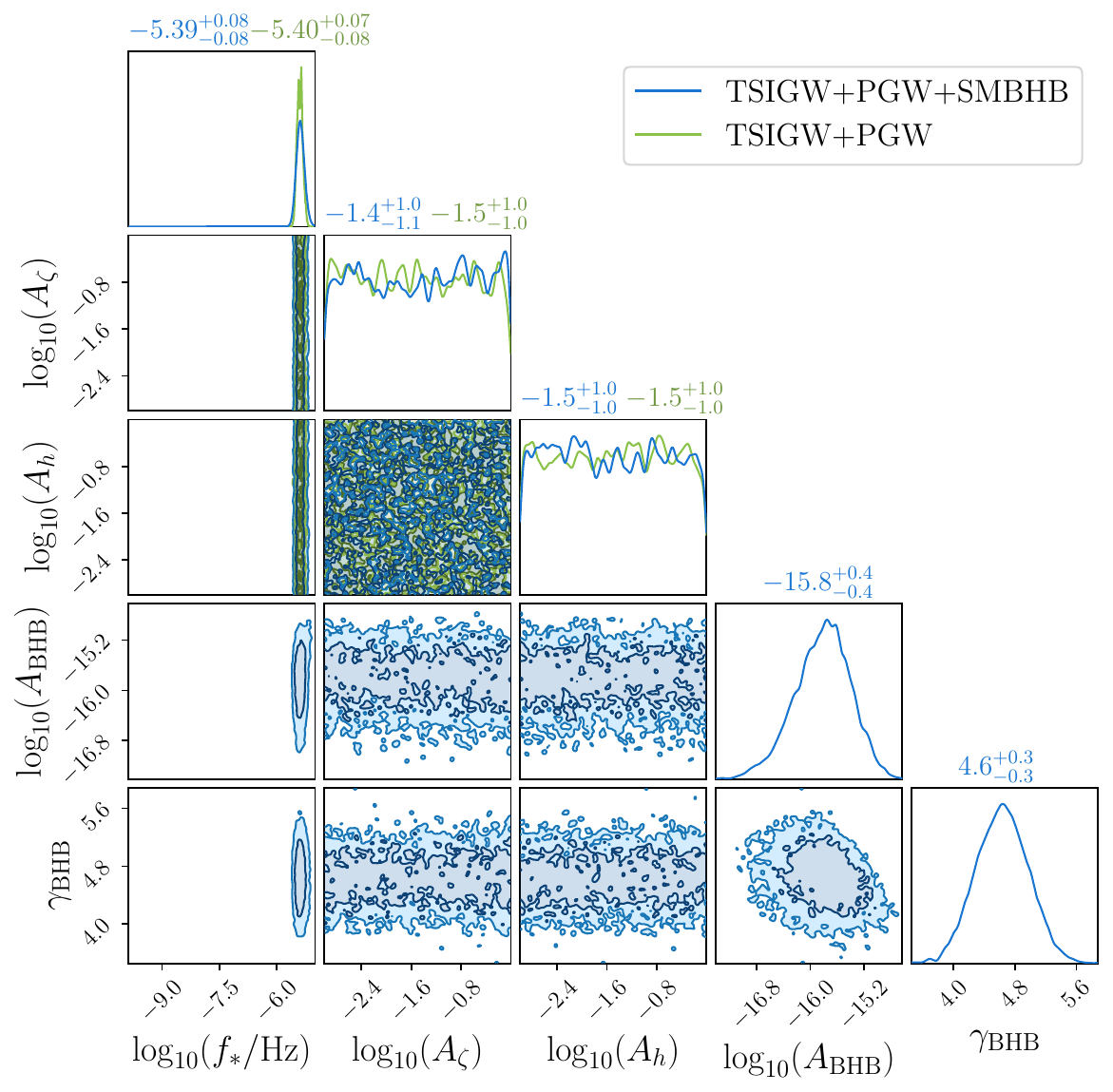}
\caption{The corner plot of the posterior distributions. The contours in the off-diagonal panels denote the $68\% $ and $95 \%$ credible intervals of the 2D posteriors. The numbers above the figures represent the median values and $1$-$\sigma$ ranges of the parameters. The blue and green solid curves correspond to the \acp{PGW}$+$\acp{TSIGW} energy spectrum with or without \ac{SMBHB}, assuming the monochromatic primordial power spectrum.} \label{fig:corner_TSIGW_mono}
\end{figure}

\begin{figure}[htbp]
    \centering
    \includegraphics[width=\columnwidth]{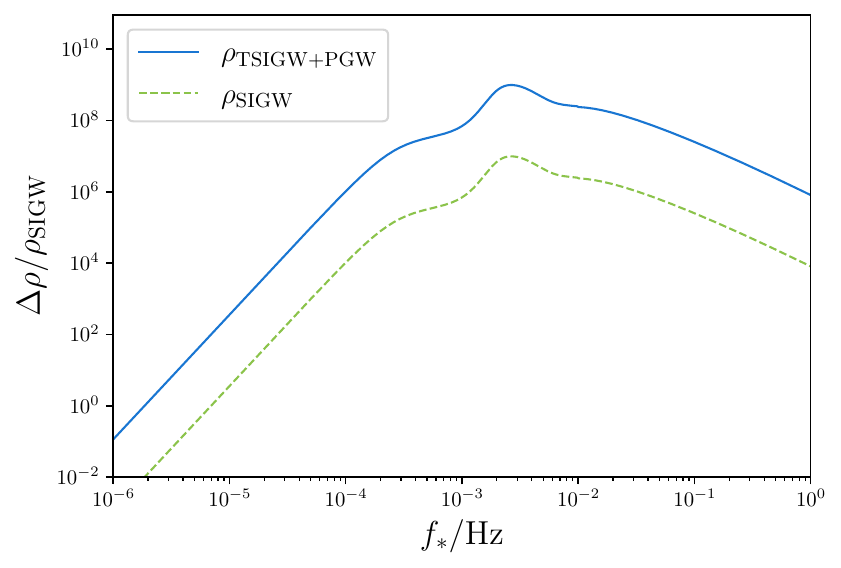}
\caption{\label{fig:SNR1d_tunefstar_mono} The \ac{SNR} of \ac{LISA} as a function of $f_*$, assuming fixed values $A_\zeta = 0.1$ and $A_h = 0.02$. The blue solid curve and green dashed curve represents the \ac{SNR} for the \acp{SIGW} and \acp{PGW}$+$\acp{TSIGW}, assuming the  monochromatic primordial power spectrum.
}
\end{figure}

We first consider the scenario where primordial curvature perturbations and primordial tensor perturbations generate a monochromatic peak on the same scale. The corresponding primordial power spectrum can be expressed as
\begin{equation}\label{eq:sd}
\begin{aligned}
    \mathcal{P}^{\delta}_{\zeta}(k)&=A_{\zeta}k_*\delta\left( k-k_* \right) \ , 
    \end{aligned}
\end{equation}
\begin{equation}\label{eq:hd}
\begin{aligned}
    \mathcal{P}^{\delta}_{h}(k)&=A_{h}k_*\delta\left(k-k_*  \right) \ ,
    \end{aligned}
\end{equation}
where $k_*$ is the wavenumber at which the power spectrum has a $\delta$ function peak. $A_{h}$ and $A_{\zeta}$ respectively denote the amplitudes of the primordial curvature perturbation and primordial tensor perturbation power spectra. By substituting Eq.~(\ref{eq:sd}) and Eq.~(\ref{eq:hd}) into Eq.~(\ref{eq:1Oss})--Eq.~(\ref{eq:1Ott}), we obtain the analytical expression for the energy density spectrum of second-order \acp{TSIGW} under a monochromatic primordial power spectrum, given by
\begin{equation}\label{eq:anss}
\begin{aligned}
\Omega_{\mathrm{GW},\delta}^{ss}(\tilde{k})&=A_{\zeta}^2
\frac{3}{1024}\,(4-\tilde{k}^{2})^{2}\tilde{k}^{2}\bigl(3\tilde{k}^{2}-2\bigr)^{2} \\
&\Bigl[
  \left(\bigl(3\tilde{k}^{2}-2\bigr)\,
  \ln\!\left|1-\frac{4}{3\tilde{k}^{2}}\right|+4\right)^{2} \\
& +\pi^{2}\bigl(3\tilde{k}^{2}-2\bigr)^{2}\,
  \Theta(2\sqrt{3}-3\tilde{k})
\Bigr] \Theta(2-\tilde{k})\ ,
\end{aligned}
\end{equation}

\begin{equation}\label{eq:anst}
\begin{aligned}
\Omega_{\mathrm{GW},\delta}^{st}(\tilde{k})&=A_h A_{\zeta}
\frac{\tilde{k}^{2}}{442\,368}(16+24\tilde{k}^{2}+\tilde{k}^{4})\\
&\Bigl[
  3\bigl(4-24\tilde{k}^{2}+9\tilde{k}^{4}\bigr)^{2}\pi^{2}\Theta\!\Bigl(-3-\tfrac{2}{\tilde{k}^{2}}+\tfrac{6}{\tilde{k}}\Bigr)\, \\
&  +\Bigl(
    4\!\bigl(9\tilde{k}^{2}-8\bigr) \\
&    +\sqrt{3}\bigl(4-24\tilde{k}^{2}+9\tilde{k}^{4}\bigr)
\ln\!\Bigl|\tfrac{4+2\sqrt{3}-3\tilde{k}^{2}}{4-2\sqrt{3}-3\tilde{k}^{2}}\Bigr|
  \Bigr)^{2} \\
& \Bigr]  \Theta(2-\tilde{k}) \ ,
\end{aligned}
\end{equation}

\begin{equation}\label{eq:antt}
\begin{aligned}
\Omega_{\mathrm{GW},\delta}^{tt}(\tilde{k})&=A_{h}^2
\frac{\tilde{k}^{2}(\tilde{k}^{2}-4)^{2}}{315\,728}\,
\Bigl[ 56\tilde{k}^{6}\pi^{2}+\tilde{k}^{8}\pi^{2}  \\
 &+128\bigl(4+\pi^{2}\bigr) +128\tilde{k}^{2}\bigl(6+7\pi^{2}\bigr)  \\
 &+16\tilde{k}^{4}\bigl(4+35\pi^{2}\bigr) \\
 &+16\bigl(32+120\tilde{k}^{2}+30\tilde{k}^{4}+\tilde{k}^{6}\bigr)
      \ln\!\left|1-\tfrac{4}{\tilde{k}^{2}}\right|\\[2pt]
&+\bigl(128+896\tilde{k}^{2}+560\tilde{k}^{4}+56\tilde{k}^{6}+\tilde{k}^{8}\bigr) \\
 &     \ln^{2}\!\left|1-\tfrac{4}{\tilde{k}^{2}}\right|
\Bigr]\Theta(2-\tilde{k}) \ ,
\end{aligned}
\end{equation}
where $\tilde{k}=k/k_*$.  {Fig.~\ref{fig:spectrum_mono} illustrates the energy density spectra of second-order \acp{GW} arising from the three source terms of \acp{TSIGW} under a monochromatic primordial power spectrum, along with the corresponding \acp{PGW} spectrum. As indicated, when substantial primordial curvature perturbations coexist with primordial tensor perturbations at small scales, the tensor-tensor and tensor-scalar induced contributions produce notable corrections to the high-frequency energy density spectrum of second-order \acp{SIGW}. In this case, the total \acp{TSIGW}$+$\acp{PGW} spectrum, represented by the four curves in Fig.~\ref{fig:spectrum_mono}, should be considered.} Fig.~\ref{fig:violinplot} presents the energy density spectra of second-order \acp{SIGW} and \acp{TSIGW}, while the corresponding posterior distributions constrained by \ac{PTA} observations are shown in Fig.~\ref{fig:corner_SIGW_mono} and Fig.~\ref{fig:corner_TSIGW_mono}. The prior distributions of $\log_{10}(A_{\zeta})$, $\log_{10}(A_{h})$, and $\log_{10}(f_*/\mathrm{Hz})$ are set as uniform distributions over the intervals $[-3,0]$, $[-3,0]$, and $[-10,-5]$, respectively. To better investigate the physical characteristics of the energy density spectrum of \acp{TSIGW} across different frequency bands, we present the \ac{SNR} of \ac{LISA} for various peak scales of the primordial power spectrum in Fig.~\ref{fig:SNR1d_tunefstar_mono}. The results indicate that \acp{TSIGW} can only be significantly detected by LISA when the frequency $f_*$ exceeds $10^{-4}$ Hz. This implies that \acp{TSIGW} cannot be observed in the \ac{LISA} band if they are to dominate the \ac{PTA} signal. This feature sharply contrasts with astrophysical \ac{SGWB} and provides a potential criterion to distinguish between the two types of \ac{SGWB} \cite{Ellis:2023oxs}.


\subsection{Log-normal spectra}\label{sec:4.2}
\begin{figure}[htbp]
\centering
\includegraphics[width=\linewidth]{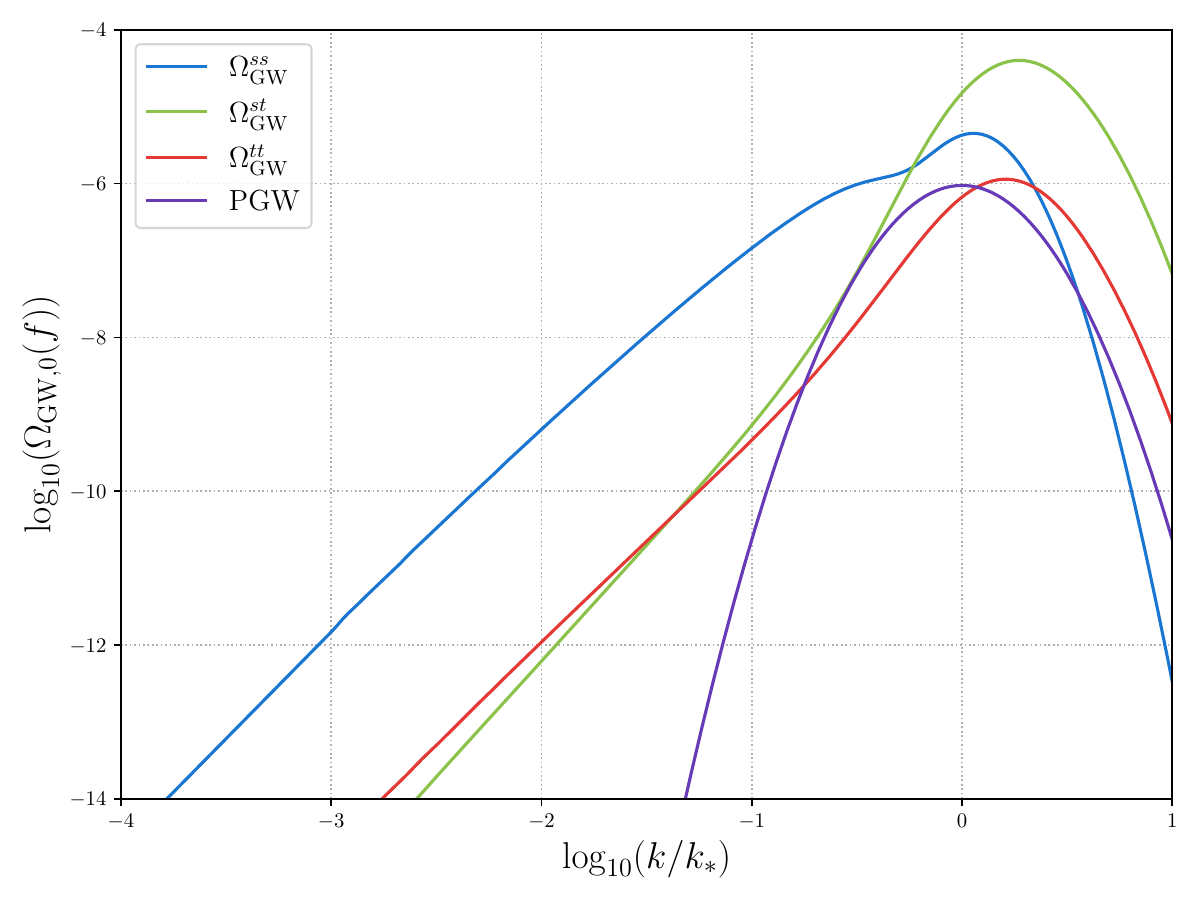}
\caption{In the case of the \ac{LN} primordial power spectrum, the current energy density spectra of the three components of second-order \acp{TSIGW} are shown as blue, green, and red curves, respectively. The corresponding energy density spectrum of \acp{PGW} is represented by the purple curve. The parameters for the curves are $A_\zeta=1$, $A_h=1$ and $\sigma=0.5$.} \label{fig:spectrum_logn}
\end{figure}

\begin{figure}[htbp]
\centering
\includegraphics[width=\linewidth]{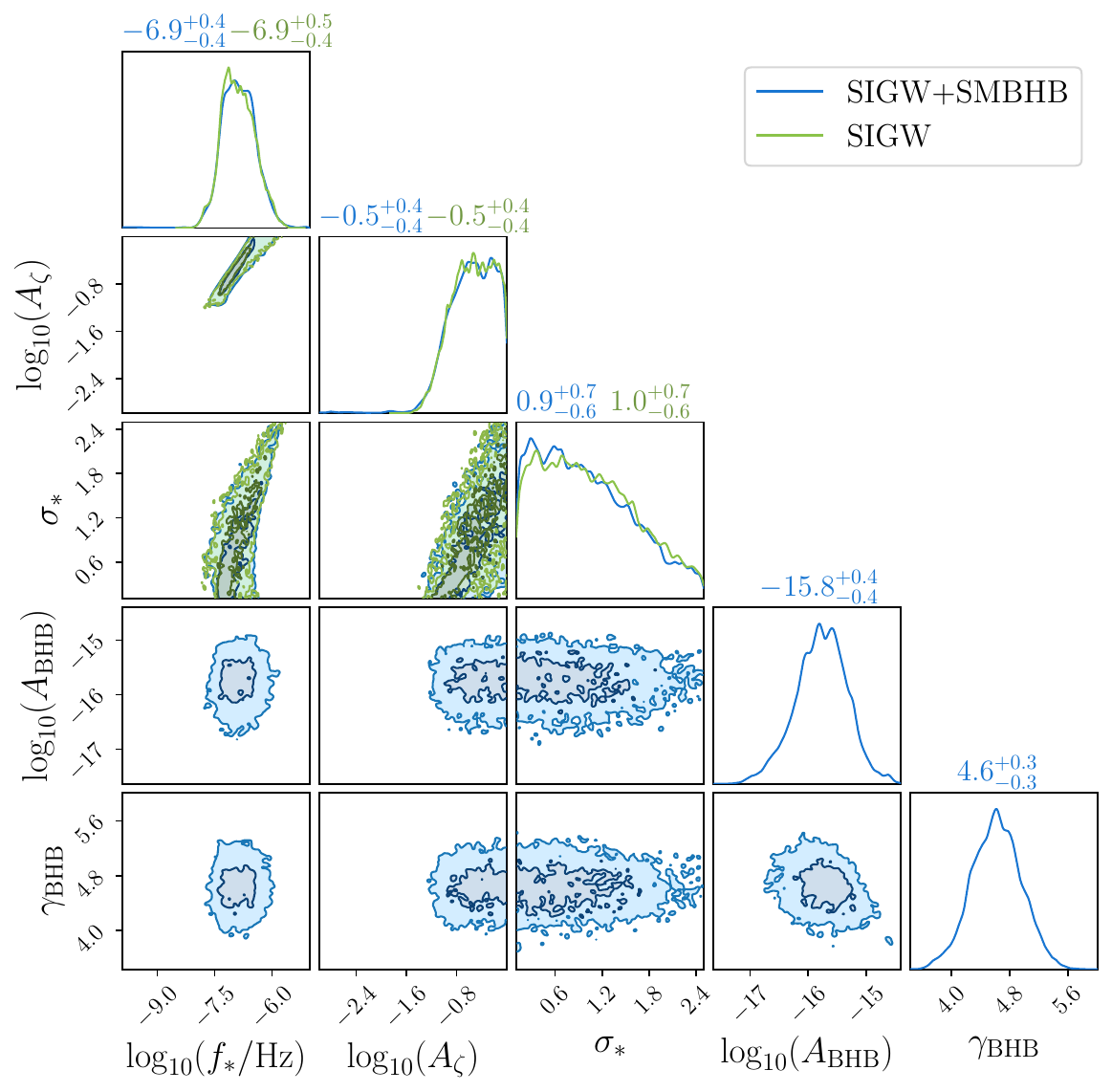}
\caption{The corner plot of the posterior distributions. The contours in the off-diagonal panels denote the $68\% $ and $95 \%$ credible intervals of the 2D posteriors. The numbers above the figures represent the median values and $1$-$\sigma$ ranges of the parameters. The blue and green solid curves correspond to the \acp{SIGW} energy spectrum with or without \ac{SMBHB}, assuming the \ac{LN} primordial power spectrum.} \label{fig:corner_SIGW_logn}
\end{figure}

\begin{figure}[htbp]
\centering
\includegraphics[width=\linewidth]{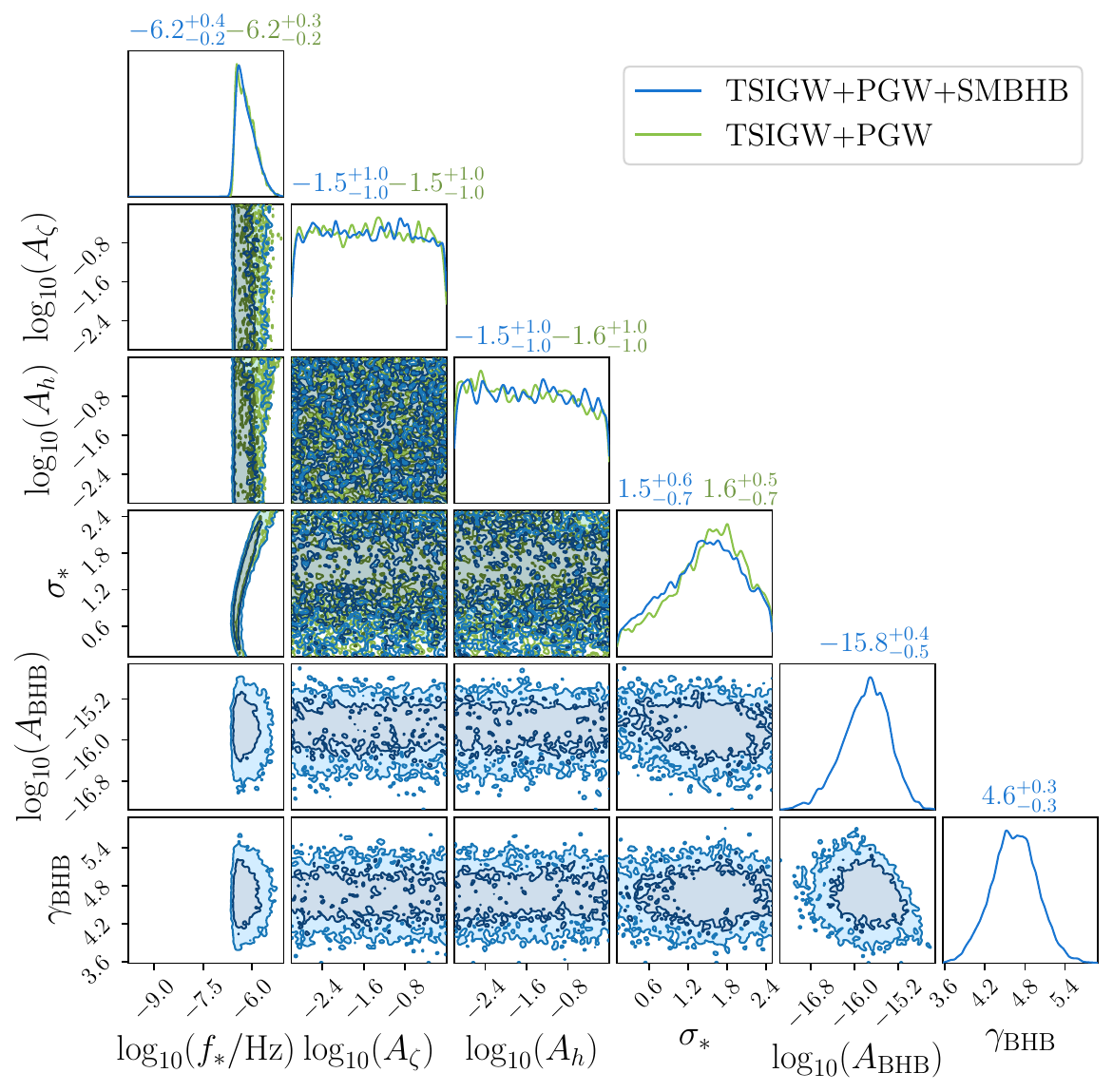}
\caption{The corner plot of the posterior distributions. The contours in the off-diagonal panels denote the $68\% $ and $95 \%$ credible intervals of the 2D posteriors. The numbers above the figures represent the median values and $1$-$\sigma$ ranges of the parameters. The blue and green solid curves correspond to the \acp{PGW}$+$\acp{TSIGW} energy spectrum with or without \ac{SMBHB}, assuming the \ac{LN} primordial power spectrum.} \label{fig:corner_TSIGW_logn}
\end{figure}

\begin{figure}[htbp]
    \centering
    \includegraphics[width=\columnwidth]{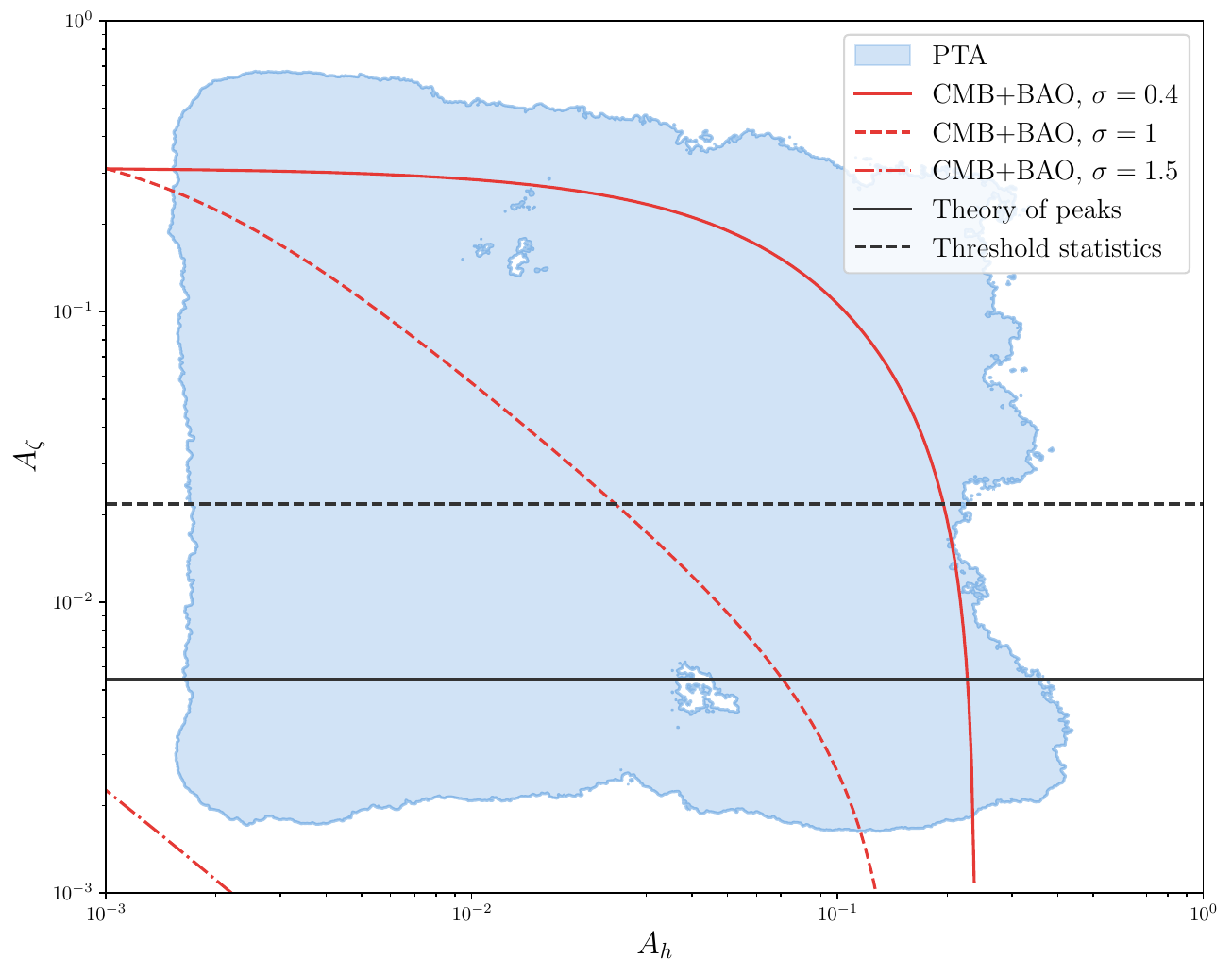}
\caption{\label{fig:constrain_LN} The constraints on  $A_\zeta$ and $A_h$ assuming the \ac{LN} primordial power spectrum. The blue shaded region represents the $68\%$ credible intervals corresponding to the two-dimensional posterior distribution shown in Fig.~\ref{fig:corner_TSIGW_logn} with the \ac{KDE} method. The red lines (solid, dashed, and dash-dotted) indicate the upper bounds on the amplitudes $A_{\zeta}$ and $A_h$ from Eq.~(\ref{eq:rhup}) with $\sigma=0.4, 1, \text{and } 1.5$ respectively. The solid black line and the dashed black line represent the upper bounds on the \ac{PBH} abundance derived from peak theory and threshold statistics with $\sigma=0.4$, respectively \cite{Carr:2020gox}.
}
\end{figure}

\begin{figure}[htbp]
    \centering
    \includegraphics[width=\columnwidth]{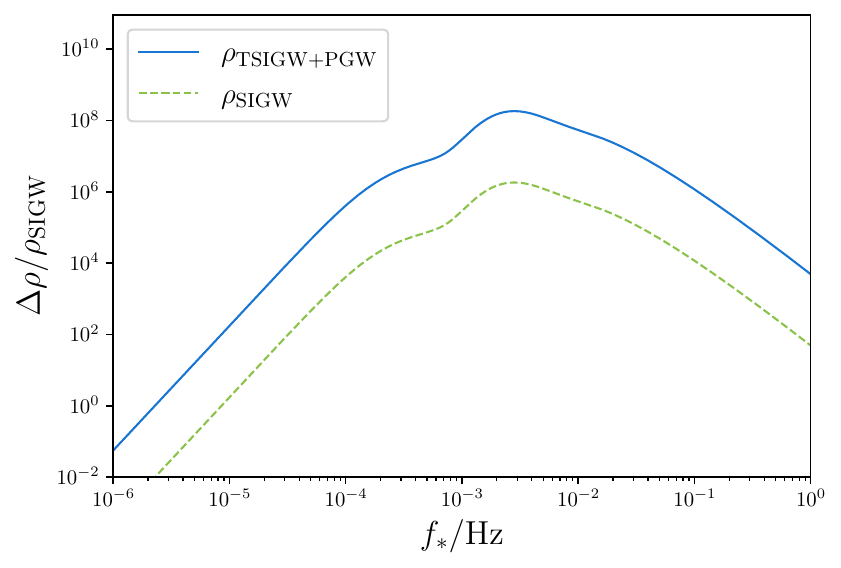}
\caption{\label{fig:SNR1d_tunefstar_LN} The \ac{SNR} of \ac{LISA} as a function of $f_*$, with fixed values $A_\zeta = 0.1$, $A_h = 0.02$ and $\sigma=0.5$. The blue solid curve and green dashed curve represents the \ac{SNR} for the \acp{SIGW} and \acp{PGW}$+$\acp{TSIGW}, assuming the \ac{LN} primordial power spectrum. 
}
\end{figure}

Since the monochromatic power spectrum exhibits an infinitely high and infinitely narrow peak, standard inflationary models generally cannot realize such a primordial power spectrum. Therefore, we consider the more realistic \ac{LN} primordial power spectra \cite{Pi:2020otn}
\begin{equation} \label{eq:LN1}
\mathcal{P}^{\mathrm{LN}}_\zeta (k)=\frac{A_{\zeta}}{\sqrt{2 \pi \sigma^2} } \exp \left(-\frac{1}{2 \sigma^2} \ln ^2\left(k / k_*\right)\right) \ ,
\end{equation}
\begin{equation} \label{eq:LN2}
\mathcal{P}^{\mathrm{LN}}_h (k)=\frac{A_{\zeta}}{\sqrt{2 \pi \sigma^2} } \exp \left(-\frac{1}{2 \sigma^2} \ln ^2\left(k / k_*\right)\right) \ ,
\end{equation}
where $A_{\zeta}$ and $A_h$ are amplitudes of primordial power spectra and $k_*=2\pi f_*$ is the wavenumber at which the primordial power spectrum has a \ac{LN} peak. The parameter $\sigma$ indicates the width of the \ac{LN} primordial power spectrum. In Fig.~\ref{fig:spectrum_logn}, we present the energy density spectra of second-order \acp{GW} generated by the three source terms of \acp{TSIGW} under the \ac{LN} primordial power spectrum, together with the corresponding \acp{PGW} spectrum. Furthermore, Fig.~\ref{fig:corner_SIGW_logn} and Fig.~\ref{fig:corner_TSIGW_logn} illustrate the posterior distributions for second-order \acp{SIGW} and second-order \acp{TSIGW}, respectively. The prior distributions of $\log_{10}(A_{\zeta})$, $\log_{10}(A_{h})$,  $\log_{10}(f_*/\mathrm{Hz})$, and $\sigma$ are set as uniform distributions over the intervals $[-3,0]$, $[-3,0]$, $[-10,-5]$, and $[0.1,2.5]$, respectively. Based on current \ac{PTA}$+$\ac{PBH}$+$\ac{CMB}$+$\ac{BAO} observations, we derive constraints on the small-scale amplitude of the primordial power spectra. As shown in Fig.~\ref{fig:constrain_LN}, the blue-shaded region represents the posterior distribution derived from \ac{PTA} observations. The red and black curves correspond to the large-scale cosmological constraints discussed in Sec.~\ref{sec:3.1} and the \ac{PBH} abundance constraints presented in Sec.~\ref{sec:3.2}, respectively. The allowed parameter space of the primordial power spectrum inferred from current cosmological observations lies beneath the curve in Fig.~\ref{fig:constrain_LN}. Furthermore, if \acp{TSIGW} are assumed to dominate the current \ac{PTA} signal, the parameters of the primordial power spectrum must reside within the blue-shaded region of Fig.~\ref{fig:constrain_LN}. Moreover, Fig.~\ref{fig:SNR1d_tunefstar_LN} shows how the \ac{SNR} of \ac{LISA} evolves with $f_*$ in the case of the \ac{LN} primordial power spectra.


As shown in Fig.~\ref{fig:corner_TSIGW_mono} and Fig.~\ref{fig:corner_TSIGW_logn} , when considering \acp{PGW} with large amplitudes on small scales, the posterior distributions exhibit strong degeneracy between $A_{\zeta}$ and $A_h$. This degeneracy arises not only from the limited precision of current \ac{PTA} data, but also from the characteristics of the artificially parametrized primordial power spectrum. More precisely, for a manually specified primordial power spectrum, its shape,  amplitude, and peak location on small scales can be adjusted through different parameters. When the shape of the primordial power spectrum is fixed, the relative contributions of the three energy density spectra described in Eq.~(\ref{eq:1Oss})--Eq.~(\ref{eq:1Ott}) can be tuned by varying the parameters $k_*$, $A_{\zeta}$, and $A_h$. Under these conditions, a certain degree of parameter degeneracy in the region constrained by PTA observations becomes inevitable. However, as will be discussed in Sec.~\ref{sec:5.0}, such degeneracies may not exist for primordial spectra derived from specific models.

\subsection{Dominant contributor to \ac{PTA}: \acp{SIGW} or \acp{TSIGW}?}\label{sec:4.3}
Based on the theoretical results of \acp{SIGW} and \acp{TSIGW}, and in combination with current \ac{PTA} observational data, we present in Fig.~\ref{fig:violinplot} the energy density spectra of induced gravitational waves under four different scenarios. The results in Fig.~\ref{fig:violinplot} indicate that, whether considering a monochromatic primordial power spectrum or a \ac{LN} primordial power spectrum, both \acp{SIGW} and \acp{TSIGW} appear capable of dominating the current \ac{PTA} observations. However, as discussed in Sec.~\ref{sec:3.0}, beyond \ac{PTA} observations, \acp{SIGW} and \acp{TSIGW} must also satisfy constraints from other cosmological observations. More precisely, the energy density spectra of \acp{SIGW} and \acp{TSIGW} must satisfy the constraints imposed by current cosmological observations such as BAO, CMB, and \ac{BBN}, which set an upper bound on the energy density of \ac{SGWB}. Refs.~\cite{Cang:2023ysz,Cang:2022jyc,Wang:2023sij} systematically investigated the constraints from these cosmological observations on second-order \acp{SIGW}, showing that \acp{SIGW} can dominate the current PTA observations without violating the corresponding cosmological bounds. 

In this study, we investigate the possibility that \acp{TSIGW} dominate the current \ac{PTA} observations, as well as the constraints imposed on such waves by \ac{CMB}$+$\ac{BAO} data. As shown in Fig.~\ref{fig:constrain_LN},  the region below the red curves corresponds to the parameter space allowed by \ac{CMB}$+$\ac{BAO} observations given in Eq.~(\ref{eq:rhup}), while the blue area represents the posterior distribution derived from \ac{PTA} observations. For a narrow \ac{LN} primordial power spectrum width $\sigma$, \acp{TSIGW} can both satisfy \ac{CMB}$+$\ac{BAO} bounds and dominate PTA signals. Conversely, a broad spectrum width leads to excessive high-frequency amplitudes in the energy density spectrum induced by tensor-scalar source terms, conflicting with \ac{CMB}$+$\ac{BAO} constraints. A similar issue may arise when the peak of the primordial curvature perturbation power spectrum and that of the primordial tensor perturbation spectrum lie in different frequency bands, leading to excessive high-frequency energy density from tensor-scalar source terms. Preliminary investigations of this problem have been reported in Refs.~\cite{Yu:2023lmo,Bari:2023rcw,Picard:2023sbz}, and we will further discuss it in Sec.~\ref{sec:4.4}.

As discussed in Sec.~\ref{sec:3.2}, beyond the cosmological constraints discussed above, the upper limit on \ac{PBH} abundance further restricts the parameter space of the power spectrum of primordial curvature perturbation, indirectly influencing the energy density spectrum of induced gravitational waves. For \acp{SIGW}, Ref.~\cite{Franciolini:2023pbf} provided detailed discussions on this issue. The results indicate that if \acp{SIGW} dominate the current \ac{PTA} observations, they would lead to an overproduction of \ac{PBH}, thereby violating the existing \ac{PBH} observational constraints. Consequently, the second-order \acp{SIGW} calculated from Eq.~(\ref{eq:1Oss}) cannot dominate the \ac{PTA} observations. By introducing primordial non-Gaussianity or a variable sound speed, the tension between \ac{PBH} overproduction and \acp{SIGW} dominating \ac{PTA} observations can be partially alleviated \cite{Balaji:2023ehk,Liu:2023ymk,Wang:2023ost,Adshead:2021hnm}. In contrast, \acp{TSIGW} can dominate \ac{PTA} observations without \ac{PBH} overproduction. Specifically, as shown in Fig.~\ref{fig:constrain_LN}, the parameter space constraints obtained from two different \ac{PBH} abundance calculation methods lie below the black solid and dashed lines, respectively. Within this parameter space, \acp{TSIGW} can still dominate \ac{PTA} observations. The presence of large primordial tensor perturbations at small scales reduces the required amplitude of primordial curvature perturbations for \acp{TSIGW} to dominate \ac{PTA} observations, thereby avoiding the \ac{PBH} overproduction problem.

\subsection{Primordial power spectra with peaks located at different scales}\label{sec:4.4}
Second-order \acp{GW} induced by tensor-scalar source terms exhibit potential divergence issues under certain conditions. As mentioned in Refs.~\cite{Picard:2023sbz,Bari:2023rcw}, Eq.~(\ref{eq:1Ost}) may diverge with $v^{-4}$ when $v\to 0$ and $u\to 1$. Similarly, it may diverge with $u^{-4}$ when $u\to 0$ and $v\to 1$. Ref.~\cite{Bari:2023rcw} explores the origin of such divergences and provides a phenomenological approach to handle them. Specifically, as shown in Eq.~(\ref{eq:h2}), $u$ and $v$ respectively denote the wave numbers associated with first-order scalar and tensor perturbations. When the wavelength of the first-order scalar perturbation significantly exceeds that of the first-order tensor perturbation, the first-order scalar perturbation can no longer be regarded as a localized source term. Conversely, when the wavelength of the first-order tensor perturbation far exceeds that of the scalar perturbation, the tensor perturbation similarly ceases to serve as a localized source term.

To eliminate this potential divergence, Ref.~\cite{Bari:2023rcw} proposed the artificial insertion of two functions into Eq.~(\ref{eq:1Ost})
\begin{equation}\label{eq:fu}
    f(u)=\frac{u^4}{d^4+u^4}  \ , \  f(v)=\frac{v^4}{d^4+v^4} \ ,
\end{equation}
where $d\approx \mathcal{O}(0.1) \sim \mathcal{O}(1)$. As shown in Eq.~(\ref{eq:fu}), $f(x)$ goes from $f(x\gg 1)\approx 1$ to $f(x\ll 1)\approx (x/d)^4$ and cures any divergence. By inserting Eq.~(\ref{eq:fu}) into Eq.~(\ref{eq:1Ost}), we can effectively eliminate the potential "divergence" issue in Eq.~(\ref{eq:1Ost}).

\begin{figure}[htbp]
\centering
\includegraphics[width=\linewidth]{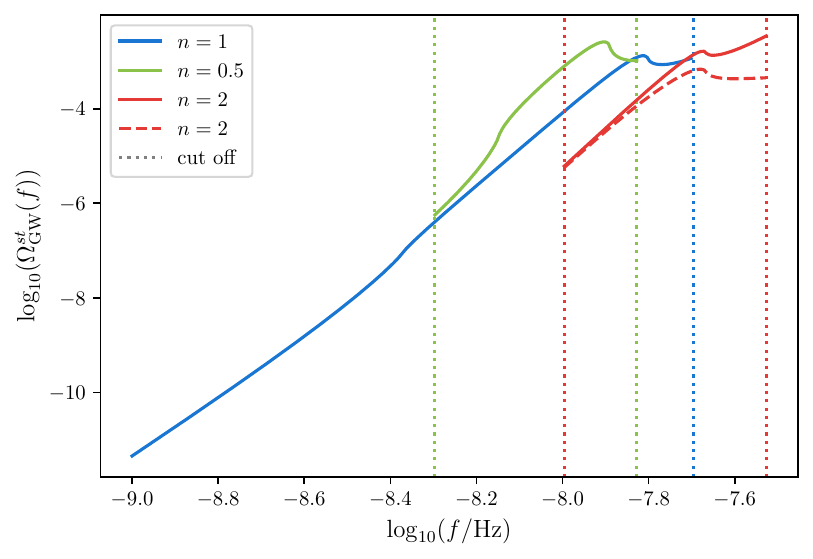}
\caption{The energy density spectra of second-order gravitational waves induced by the tensor-scalar source term during \ac{RD} era. The parameters for the curves are $A_\zeta=0.1$, $A_h=0.1$, ${f_{\zeta}*}=10^{-8}$Hz, $n={{f_{h}*}}/{{f_{\zeta}*}}$. The solid curves and dashed curve represent the the energy density spectra with or without functions in Eq.~(\ref{eq:fu}). The dotted lines represent the physical cut off. } \label{fig:spectrum_tsigw}
\end{figure}

For cases where the peak scales of primordial tensor perturbations and scalar perturbations differ, we consider the following form of the monochromatic primordial power spectra
\begin{equation}\label{eq:de1n}
\begin{aligned}
    \mathcal{P}^{\delta}_{\zeta}(k)&=A_{\zeta}k_{\zeta*}\delta\left( k-k_{\zeta*} \right) \ , 
    \end{aligned}
\end{equation}
\begin{equation}\label{eq:de2n}
\begin{aligned}
    \mathcal{P}^{\delta}_{h}(k)=A_{h}k_{h*}\delta\left(k-k_{h*}  \right) \ .
    \end{aligned}
\end{equation}
Here, we define the parameter $n=k_{h*}/k_{\zeta*}$ characterizing the relative position of the peaks in the two primordial power spectra. When $n=1$, the peaks of both spectra coincide. By rewriting Eq.~(\ref{eq:1Ost}) as
\begin{equation}\label{eq:Rwr}
\begin{aligned}
    \Omega_{\mathrm{GW}}^{st}(k)&=\int_{0}^{\infty} \mathrm{d}v \int_{|1-v|}^{1+v} \mathrm{d}u\, \mathcal{P}_{\zeta}(uk) \mathcal{P}_h(vk)  \\
        &\times \frac{1}{442368 u^8 v^8} \mathcal{T}^{st}\left(u,v  \right) \ ,
 \end{aligned}
\end{equation}
the corresponding analytical expression for the energy density spectrum from the primordial power spectra in Eq.~(\ref{eq:de1n}) and Eq.~(\ref{eq:de2n}) is given by
\begin{equation}\label{eq:adnst}
\begin{aligned}
&\Omega_{\mathrm{GW}}^{st}(\tilde{k}_{h},\tilde{k}_{\zeta})= A_{\zeta}A_{h}\frac{\tilde{k}_{\zeta}^{7}\tilde{k}_{h}^{7}}{442\,368}\mathcal{T}^{st}\left(\frac{1}{\tilde{k}_{\zeta}},\frac{1}{\tilde{k}_{h}} \right) \\
&~~\times\Theta(\lvert 1+\tfrac{1}{\tilde{k}_{h}} \lvert -\tfrac{1}{\tilde{k}_{\zeta}}) \Theta(\tfrac{1}{\tilde{k}_{\zeta}}-\lvert 1-\tfrac{1}{\tilde{k}_{h}} \lvert)  \ ,
\end{aligned}
\end{equation}
where $\tilde{k}_{h}=k/k_{h*}=k/(nk_{\zeta*})=\tilde{k}_{\zeta}/n$. Since the shape and amplitude of the energy density spectrum $\Omega_{\mathrm{GW}}^{ss}$ generated by scalar-scalar source terms and spectrum $\Omega_{\mathrm{GW}}^{tt}$ generated by tensor-tensor source terms remain unchanged, we focus on the second-order gravitational waves $\Omega_{\mathrm{GW}}^{st}$ produced by tensor-scalar source terms. In Fig.~\ref{fig:spectrum_tsigw}, we present the energy density spectra of second-order gravitational waves induced by tensor-scalar source terms $\Omega_{\mathrm{GW}}^{st}$ the during \ac{RD} era.

As shown in Eq.~(\ref{eq:adnst}) and Fig.~\ref{fig:spectrum_tsigw}, when considering monochromatic primordial power spectra $\mathcal{P}_{\zeta}(k)$ and $\mathcal{P}_{h}(k)$ located at different scales, the energy density spectrum $\Omega_{\mathrm{GW}}^{st}$ exhibits a distinct cutoff. Specifically, when $n>1$, the peak scale of the primordial gravitational waves is smaller than that of the primordial curvature perturbations. In this case, the non-zero region of the energy density spectrum $\Omega_{\mathrm{GW}}^{st}$ lies within $n-1<\tilde{k}_{\zeta}<n+1$. Conversely, when $n<1$, the peak scale of the primordial gravitational waves exceeds that of the primordial curvature perturbations, and the non-zero region is $1-n<\tilde{k}_{\zeta}<1+n$. As $n$ approaches zero, the non-zero region of $\Omega_{\mathrm{GW}}^{st}$ becomes increasingly narrow. Moreover, when the peak scales of the primordial gravitational waves and primordial curvature perturbations differ, the amplitude of the energy density spectrum  $\Omega_{\mathrm{GW}}^{st}$ increases significantly. This amplification effect becomes increasingly pronounced as the discrepancy between the scales of the primordial power spectra $\mathcal{P}_{\zeta}(k)$ and $\mathcal{P}_{h}(k)$ grows \cite{Yu:2023lmo}. Fig.~\ref{fig:spectrum_tsigw} compares the energy density spectrum $\Omega_{\mathrm{GW}}^{st}$ before and after the insertion of function $f(x)$, as depicted by the solid and dashed lines. It should be emphasized that the method we present here is purely phenomenological and aims to suppress the energy density spectrum $\Omega_{\mathrm{GW}}^{st}$. The authenticity of this spectrum amplification effect and the development of a rigorous first-principles approach to the treatment of \acp{TSIGW} under these conditions have yet to be definitively resolved \cite{Bari:2023rcw}.

\section{ Tensor induced gravitational waves }\label{sec:5.0}
In previous sections, we considered second-order \acp{GW} induced by large-amplitude primordial curvature perturbations and primordial tensor perturbations at small scales. In this section, we focus on the scenario in which only large primordial tensor perturbations exist at small scales, without the presence of significant primordial curvature perturbations. We refer to the higher-order \acp{GW} induced solely by first-order tensor perturbations as \acp{TIGW}. Unlike \acp{SIGW}, \acp{TIGW} remain gauge-independent due to the intrinsic gauge invariance of first-order tensor perturbations \cite{Hwang:2017oxa,Inomata:2019yww,Lu:2020diy,Domenech:2020xin,Tomikawa:2019tvi,Ota:2021fdv}. Disregarding the source terms $\mathcal{S}^{(2)}_{lm,\phi\phi}$ and $\mathcal{S}^{(2)}_{lm,\phi h}$ in Eq.~(\ref{eq:h}) yields the evolution equation for second-order \acp{TIGW}, with their energy density spectrum described by Eq~(\ref{eq:1Ott}). 

In this section, we examine a specific inflationary model, the Nieh-Yan modified Teleparallel Gravity model, which is capable of generating large-amplitude primordial gravitational waves at small scales \cite{Fu:2023aab,Cai:2021uup}. The Nieh-Yan modified Teleparallel Gravity model is characterized by the following action \cite{Li:2020xjt,Li:2021wij}
\begin{equation}\label{eq:TTT}
\begin{aligned}
S= & \int d^4 x \sqrt{-g}\left[-\frac{R}{2}+\frac{\alpha \phi}{4} \mathcal{T}_{A \mu \nu} \tilde{\mathcal{T}}^{A \mu \nu}\right. \\
&\left.+\frac{1}{2} \nabla_\mu \phi \nabla^\mu \phi-V(\phi)\right] +S_{\text {other }} \ .
\end{aligned}
\end{equation}
Here, the $\phi$ field emerges as a dynamical scalar field, and the action $S_{\text {other }}$ describes an additional canonical field that is minimally coupled to gravity. In Eq.~(\ref{eq:TTT}), $R$ is the Ricci scalar, $\tilde{\mathcal{T}}^{A \mu \nu}=(1 / 2) \varepsilon^{\mu \nu \rho \sigma} \mathcal{T}^A{ }_{\rho \sigma}$ represents the dual of the torsion two form $\mathcal{T}^A{ }_{\mu \nu}$ with $\varepsilon^{\mu \nu \rho \sigma}$ being the Levi-Civita tensor, and $\alpha$ is the coupling constant. In Nieh-Yan modified Teleparallel Gravity model with the spatially flat \ac{FLRW} metric, the background evolution is identical to that in general relativity.

Within this framework, primordial tensor perturbations display velocity birefringence and adhere to the specified equation of motion in momentum space
\begin{equation}
    h_k^{\lambda\prime\prime} + 2\mathcal{H} h_k^{\lambda\prime} + (k^2 + k\alpha A_\lambda\phi')h = 0 \ ,
\end{equation}
where $\lambda=R(L)$ denotes the right(left)-handed polarization with $A_R=1$ and $A_L=-1$. During inflation, if certain modes undergo a tachyonic instability at sub-horizon scale when the effective mass $\omega^2 = k^2 + k\alpha\lambda\phi'$ becomes negative, the power spectrum would be amplified. As a concrete example, we choose the Starobinsky linear potential \cite{Starobinsky:1992ts}
\begin{equation}
    V(\phi) = \begin{cases}
        V_0 + A_+(\phi - \phi_0) \ , & \qquad \phi > \phi_0 \ , \\
        V_0 + A_-(\phi - \phi_0) \ , & \qquad \phi \leqslant\phi_0  \ .
    \end{cases}
\end{equation}
where the parameter values are chosen as $V_0 = 10^{-14}$, $A_+ = 10^{-14}$, $A_- = 10^{-15}$, $\phi_0 = 6$. The initial value of the field $\phi$ is set to $\phi = 11.32$, corresponding to the time when the CMB scale $k_{\rm CMB} = 0.05 \mathrm{Mpc}^{-1}$ exits the horizon. In addition, the range of the parameter $\alpha$ is chosen as $\alpha\in[24, 30]$.

\begin{figure}[htbp]
\centering
\includegraphics[width=\linewidth]{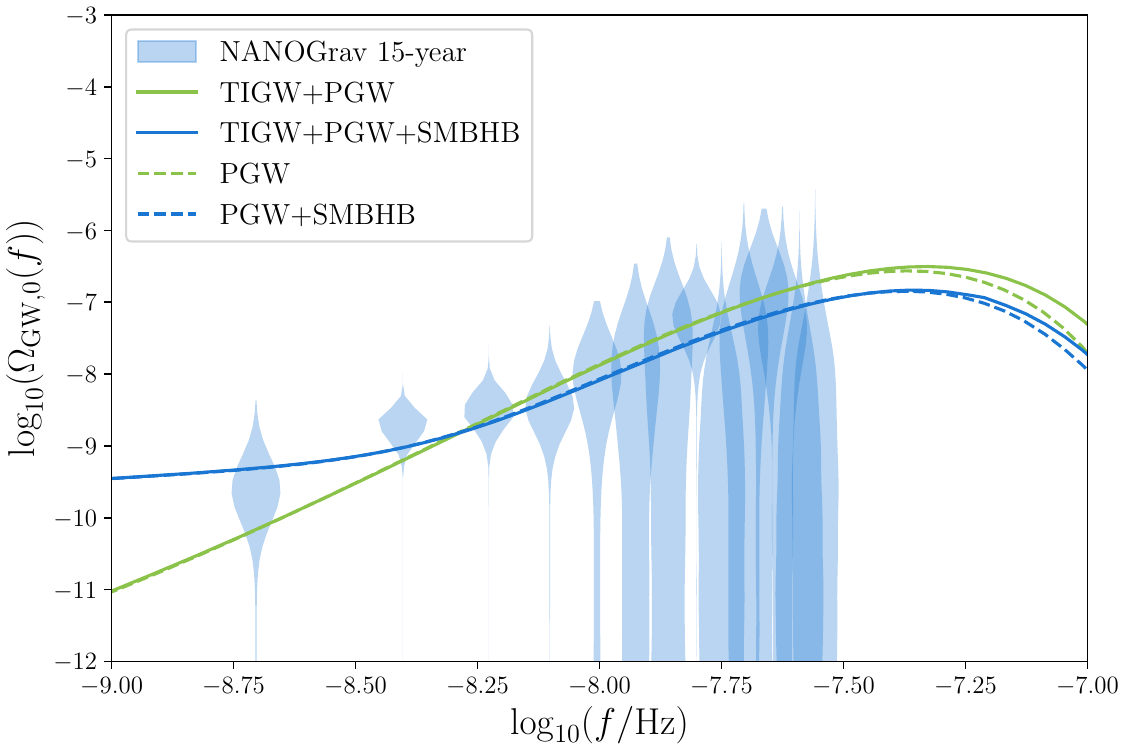}
\caption{The current energy density spectra of \acp{PGW} and \acp{PGW}$+$\acp{TIGW} with or without \ac{SMBHB}. The energy density spectra derived from the free spectrum of the NANOGrav 15-year are shown in blue. The blue and green curves represent the energy density spectra of \acp{GW} with different line styles labeled in the figure. These parameters are selected based on the median values of the posterior distributions, with the median values shown as blue and green numbers in Fig.~\ref{fig:corner_PGW_NY} and Fig.~\ref{fig:corner_TIGW_NY}.} \label{fig:violinplot_NY}
\end{figure}

\begin{figure}[htbp]
\centering
\includegraphics[width=\linewidth]{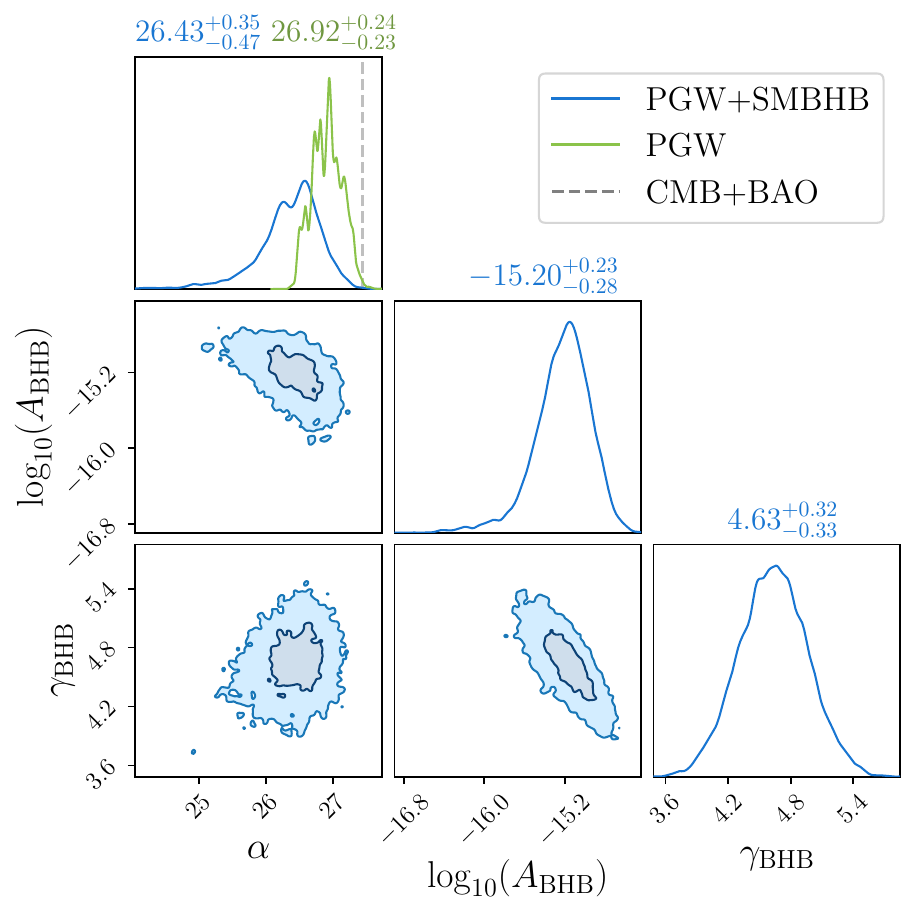}
\caption{The corner plot of the posterior distributions. The contours in the off-diagonal panels denote the $68\% $ and $95 \%$ credible intervals of the 2D posteriors. The numbers above the figures represent the median values and $1$-$\sigma$ ranges of the parameters. The blue and green solid curves correspond to the \acp{PGW} energy spectrum with or without \ac{SMBHB}. The grey dashed lines denote the upper bounds from \ac{CMB} and \ac{BAO} observations in Eq.~(\ref{eq:rhup}).} \label{fig:corner_PGW_NY}
\end{figure}
\begin{figure}[htbp]
\centering
\includegraphics[width=\linewidth]{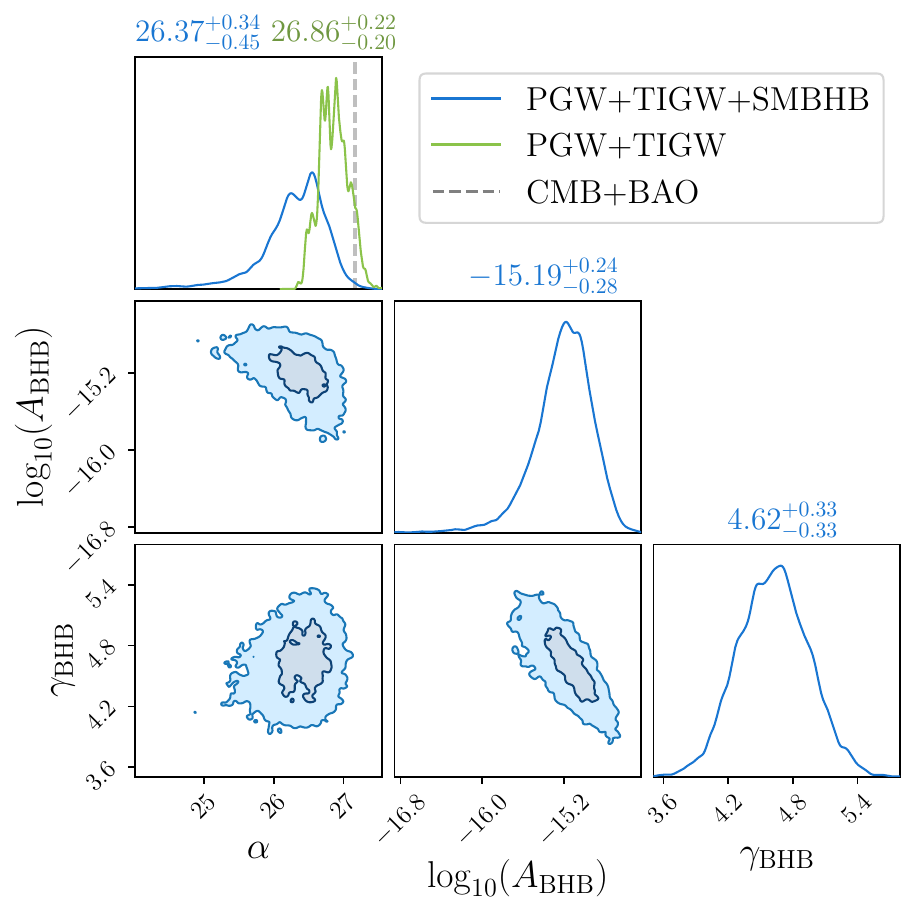}
\caption{The corner plot of the posterior distributions. The contours in the off-diagonal panels denote the $68\% $ and $95 \%$ credible intervals of the 2D posteriors. The numbers above the figures represent the median values and $1$-$\sigma$ ranges of the parameters. The blue and green solid curves correspond to the \acp{PGW}$+$\acp{TIGW} energy spectrum with or without \ac{SMBHB}. The grey dashed lines denote the upper bounds from \ac{CMB} and \ac{BAO} observations in Eq.~(\ref{eq:rhup}).} \label{fig:corner_TIGW_NY}
\end{figure}

In the context of the Nieh-Yan modified Teleparallel Gravity model, the energy density spectra of \acp{PGW} and \acp{PGW}$+$\acp{TIGW} are shown in Fig.~\ref{fig:violinplot_NY}. Moreover, in Fig.~\ref{fig:corner_PGW_NY} and Fig.~\ref{fig:corner_TIGW_NY}, we present the corresponding posterior distributions. The contours in the off-diagonal panels denote the $68\% $ and $95 \%$ credible intervals of the 2D posteriors. The numbers above the figures represent the median values and $1$-$\sigma$ ranges of the parameters. The grey dashed lines in Fig.\ref{fig:corner_PGW_NY} and Fig.\ref{fig:corner_TIGW_NY} represent the upper bounds on the parameter $\alpha$ derived from \ac{CMB} and \ac{BAO} observations in Eq.~(\ref{eq:rhup}), which are $\alpha=27.44$ and $\alpha=27.16$, respectively. 

As illustrated in Fig.~\ref{fig:violinplot_NY}, the second-order corrections to the total energy density spectrum caused by the \acp{TIGW} are predominantly concentrated in the high-frequency region, exerting minimal impact on the energy density within the \ac{PTA} band. Therefore, the posterior distributions of the parameters inferred from \ac{PTA} observations, as illustrated in Fig.~\ref{fig:corner_PGW_NY} and Fig.~\ref{fig:corner_TIGW_NY}, exhibit only minor differences. Furthermore, unlike the manually specified primordial power spectra discussed in Sec.~\ref{sec:4.0}, in Nieh-Yan modified Teleparallel Gravity model, the primordial power spectra derived from concrete models depend solely on the parameters within the model's Lagrangian. Varying these parameters can simultaneously alter the amplitude, shape, and peak location of the primordial power spectrum. Under such circumstances, it is difficult to adjust one characteristic of the spectrum without affecting the others. Consequently, the degeneracy in the parameter space for the model-derived primordial power spectra might be significantly reduced compared to artificially constructed power spectra.

\begin{figure*}[htbp]
    \centering
    \includegraphics[width=1.8\columnwidth]{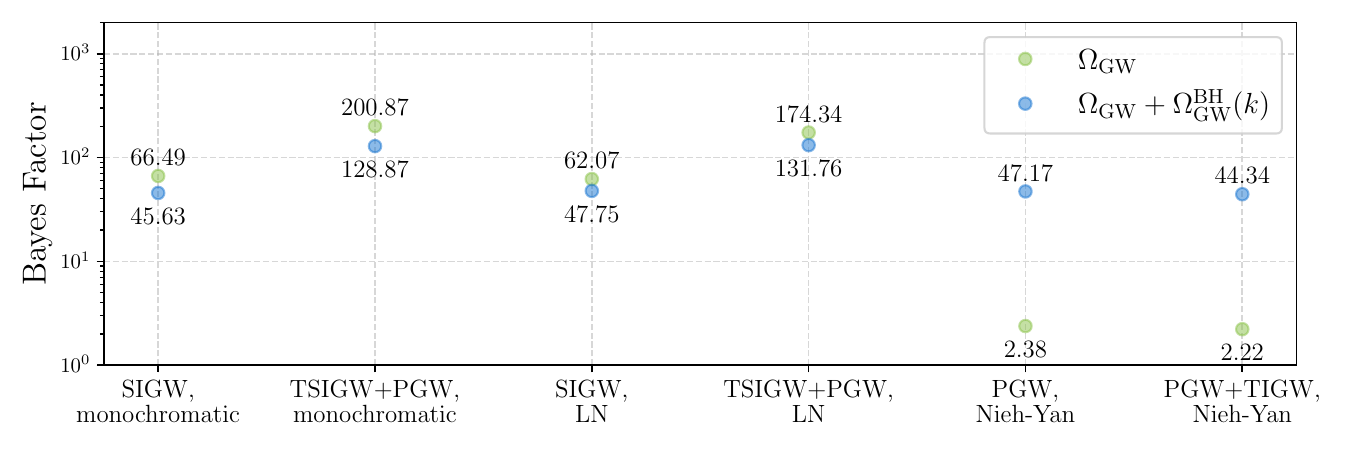}
\caption{\label{fig:bayes} The Bayes factors between different models. The vertical axis represents the Bayes factor of different models relative to \ac{SMBHB}, and the horizontal axis represents the different models. The green dots are for models without \ac{SMBHB} and the blue dots are for models in combination with the \ac{SMBHB} signal. }
\end{figure*}

As shown in Fig.~\ref{fig:bayes}, we compute the Bayes factors for different models. The results indicate that \acp{TSIGW} generated by the monochromatic primordial power spectra are more likely to dominate the current \ac{PTA} observations. It is important to note that for $n$th-order \acp{TSIGW}, the corresponding energy density spectrum $\Omega^{(n)}_{\mathrm{GW}}\sim A^n$$(A=A_\zeta,A_h)$. To ensure the convergence of cosmological perturbations, we impose conditions $A_\zeta<1$ and $A_h<1$. Consequently, in this study, we fix the upper bounds of the priors for both $A_\zeta$ and $A_h$ at $1$. Compared to Ref.~\cite{Wu:2024qdb}, we adopt a stricter upper limit on the amplitude of the primordial power spectra, which facilitates a better fit of \acp{TSIGW} to the \ac{PTA} data within the given parameter space and leads to an enhancement of the Bayes factor.

\section{Conclusion and discussion}\label{sec:6.0}
Large-amplitude primordial curvature perturbations and primordial tensor perturbations at small scales generate \acp{TSIGW} after inflation. Observations of \acp{SGWB} on varying scales allow us to constrain the physical characteristics of the primordial power spectrum on small scales. In this study, we investigated \acp{TSIGW} under various primordial power spectra and systematically analyzed the constraints imposed by current cosmological observations on the parameter space of small-scale primordial power spectra. We also examined scenarios with large primordial tensor perturbations present only on small scales and evaluated the influence of second-order \acp{TIGW} on current \ac{PTA} observations within the Nieh-Yan modified Teleparallel Gravity model. Based on current \ac{PTA} observational data, we calculated the Bayes factors for \acp{TSIGW} and \acp{TIGW} under various scenarios. Our results indicate that \acp{TSIGW} generated by the monochromatic primordial power spectra are more likely to dominate the current \ac{PTA} observations.

In this study, we investigated the influence of different forms of the primordial power spectrum on second-order \acp{TSIGW}. It should be noted that inflation models that enhance either the primordial curvature perturbation or \acp{PGW} on small scales have been extensively investigated \cite{Addazi:2024qcv,Ozsoy:2023ryl,Gorji:2023ziy,Gorji:2023sil,Jiang:2024woi,Cai:2020ovp,Addazi:2024gew,Guzzetti:2016mkm}. However, models that simultaneously produce large primordial curvature perturbations and \acp{PGW} on small scales have not yet been systematically studied. How to construct such models and how to constrain them using current cosmological observations are undoubtedly among the central topics in future research on \acp{TSIGW}. In addition to the primordial power spectra, \acp{TSIGW} may also be affected by potential new physics arising during the evolution of the universe, such as different dominant eras of the Universe \cite{Assadullahi:2009nf,Alabidi:2013lya,Domenech:2020ssp,Domenech:2020kqm,Inomata:2019zqy}, varying sound speed \cite{Balaji:2023ehk,Wright:2024awr,Chen:2024fir}, interactions between cosmological perturbations and matter \cite{Zhang:2022dgx,Yu:2024xmz,Loverde:2022wih,Saga:2014jca}, and higher-order effects \cite{Zhou:2024yke}. These related issues may be systematically explored in future work.

\vspace{0.3cm}
\begin{acknowledgements} 
This work has been funded by the National Nature Science Foundation of China under grant No. 12447127.  
\end{acknowledgements}

\bibliography{TSIGW}

\end{document}